\documentclass[a4paper,fleqn,usenatbib]{mnras}

\usepackage{mathptmx}

\usepackage[T1]{fontenc}
\usepackage{ae,aecompl}

\usepackage{graphicx}	
\usepackage{amsmath}	
\usepackage{caption}
\usepackage{amsfonts}
\usepackage{breqn}
\usepackage{amssymb}	
\usepackage{bm}
\usepackage{epsfig}
\usepackage{threeparttable}
\usepackage[justification=centering]{caption}
\usepackage{times}
\usepackage[T1]{fontenc}
\usepackage{aecompl}
\usepackage{float}
\usepackage{multirow}

\title[Asteroseismology of low-mass stars]{Asteroseismology of low-mass stars: the balance between partial ionisation and Coulomb interactions}

\author[Ana Brito \& Il\'idio Lopes]{
Ana Brito,$^{1,2}$\thanks{E-mail: anabrito@isg.pt (AB); ilidio.lopes@tecnico.ulisboa.pt (IL)}
Il\'idio Lopes,$^{2}$
\\
$^{1}$Departamento de Matem\'atica, Instituto Superior de Gest\~ao\\ Av. Marechal Craveiro Lopes, 1700-284, Lisboa, Portugal\\
$^{2}$Centro de Astrof\'{\i}sica e Gravita\c c\~ao  - CENTRA, Departamento de F\'{\i}sica, Instituto Superior T\'ecnico \\ IST, Universidade de Lisboa - UL, Av. Rovisco Pais 1, 1049-001 Lisboa, Portugal\\
}

\date{Accepted XXX. Received YYY; in original form ZZZ}

\pubyear{2021}

\begin{document}
\label{firstpage}
\pagerange{\pageref{firstpage}--\pageref{lastpage}}
\maketitle

\begin{abstract}
	
All cool stars with outer convective zones have the potential to exhibit stochastically excited stellar oscillations. In this work, we explore the outer layers of stars less massive than the Sun.
In particular, we have computed a set of stellar models raging from 0.4 to 0.9 $M_\odot$ with the aim at determining the impact on stellar oscillations of two physical processes occurring in the envelopes of these stars. Namely, the partial ionisation of chemical elements and the electrostatic interactions between particles in the outer layers. 
We find that alongside with partial ionisation, Coulomb effects also impact the acoustic oscillation spectrum. We confirm the well-known result that as the mass of a star decreases, the electrostatic interactions between particles become relevant.  We found that their impact on stellar oscillations increases with decreasing mass, and for the stars with the lowest masses ($M \lesssim 0.6 M_\odot$), it is shown that Coulomb effects dominate over partial ionisation processes producing a strong scatter on the acoustic modes. The influence of Coulomb interactions on the sound-speed gradient profile produces a strong oscillatory behaviour with diagnostic potential for the future. 

\end{abstract}

\begin{keywords}
asteroseismology -- stars: low-mass -- stars: interiors -- stars: oscillations 
\end{keywords}



\section{Introduction}\label{sec1}

\begin{figure*} 
	\centering
	\includegraphics[width=0.7\textwidth]{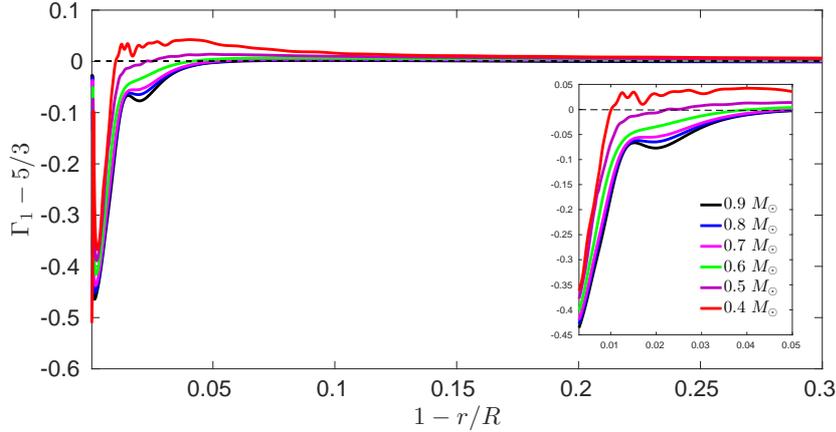}
	\caption{The first adiabatic exponent, $\Gamma_1$, plotted against the fractional radius for all the models. Variations of $\Gamma_1$ around the monatomic perfect gas value are represented in the outer layers of the stars (up to 30 per cent under the surface of the star). The inset represents the first 5 per cent of the outer region. This is a particularly important region for partial ionisation processes where helium looses its two electrons.}
	\label{fig1}
\end{figure*}

Our understanding of stellar structure and evolution has been benefiting from the enormous quantitiy of high quality observational data made available by space missions like the CoRoT \citep[e.g.][]{2007AIPC..895..201B}, \emph{Kepler/K2} \citep[e.g.][]{2010Sci...327..977B} and TESS \citep[e.g.][]{2014SPIE.9143E..20R} satellites. Asteroseismology, the study of stellar interiors by decoding the information contained in the frequencies of stellar oscillation modes is being particularly successful in the study of solar-type and red-giant stars  \citep[e.g.][]{2013ARA&A..51..353C, 2019LRSP...16....4G}. These stars have outer convective envelopes where the oscillations, known commonly as solar-like oscillations, are thought to be driven stochastically by the turbulence of near-surface convection \citep[e.g.][]{1977ApJ...212..243G}. Main-sequence stars cooler than the Sun have deep convective zones. Generally, the less massive stars have deeper convective outer regions, with the transition to a fully convective regime occurring around the value $0.35 \, M_\odot$ \citep[e.g.][]{2016ApJ...823..102C, 2018A&A...619A.177B}. All low-mass stars with outer convective zones are prone to excite solar-like oscillations. However, only a few stars with effective temperatures lesser than the Sun's effective temperature have detected solar-like oscillations. This is because these are faint stars with low luminosity values resulting in small values for the amplitudes of the oscillations. Efforts have been made to detect solar-like oscillations in low-mass stars using the high precision photometric data of the \emph{Kepler} spacecraft \citep[e.g.][]{2019FrASS...6...76R}. 

Low-mass stars are by far the most abundant type of stars in our Milky Way \citep[e.g.][]{2010AJ....139.2679B}. They are also primary targets for exoplanets' searches \citep[e.g.][]{2015ApJ...807...45D, 2018yCat..51550180M} and have a huge potential for studies related to different aspects, not yet explained, of stellar structure and evolution \citep[e.g.][]{2011ApJ...728...48K, 2013ApJ...779..183F}. Specifically, it is well known that theoretical models of low-mass stars reveal discrepancies between theoretical and observed values for several quantities such as effective temperatures or radii \citep[e.g.][]{2013ApJ...776...87S, 2014MNRAS.444.2525C}. 
Asteroseismology is an ideal tool to contribute to a better understanding of these problems. The quest for solar-like oscillations using photometric techniques in these stars is an ongoing work with missions such as TESS, and, in the near future, also the PLATO mission  \citep[e.g.][]{2014ExA....38..249R} will add to these efforts. Very promising is the use of ground-based radial velocities networks like the SONG--Stellar Observations Network Group \citep{2017ApJ...836..142G} or high precision telescopes as the KPF--Keck Planet Finder \citep{2016SPIE.9908E..70G}, which can reach high signal-to-noise ratio data in cool stars. 

In this work, we will explore, from a theoretical point of view, the outer convective envelopes of stars less massive than the Sun. To that end, we computed six stellar models raging from $0.4 $ to   $0.9 \, M_\odot$. The models were built using the MESA stellar evolution code \citep{2011ApJS..192....3P, 2013ApJS..208....4P, 2015ApJS..220...15P, 2018ApJS..234...34P, 2019ApJS..243...10P}. In particular, we will study the effect of the most relevant partial ionisation processes on the acoustic spectrum of oscillations. It is well known that in the Sun and other main-sequence solar-type stars, the partial ionisation of chemical elements, specially the second helium ionisation produces a distinct oscillatory signature not only in the frequencies but also in several seismic parameters \citep[e.g.][]{1990LNP...367..283G, 2007MNRAS.375..861H, 2014ApJ...782...18M, 2017ApJ...837...47V, 2018ApJ...853..183B}. It is also known that in cool stars, the electrostatic interactions between particles become relevant given the higher values of density in these stars. These electrostatic interactions introduce modifications to the gas pressure, leading to non-negligible deviations from the ideal gas law \citep[e.g.][]{1998asa..book.....R, 2009pfer.book.....M}.

We will study, for the first time, the impact of these two non-ideal physical processes, partial ionisation of chemical elements and electrostatic interactions between particles, on the theoretical spectrum of acoustic oscillations obtained from the model parameters. For this purpose, we use a seismic diagnostic appropriate to probe the outer convective layers of low-mass stars \citep[e.g.][]{1987SvAL...13..179B, 1989SvAL...15...27B, 1997ApJ...480..794L, 2001MNRAS.322..473L, 2017ApJ...843...75B, 2018ApJ...853..183B}. We find that these two processes compete in importance, with Coulomb effects  becoming the dominant process that impacts the oscillations for stars less massive than approximately $0.6 \, M_\odot$.

This paper is structured as follows. In Section 2, we describe the adiabatically stratified convective regions of the stellar models. Namely, we will focus on some of its thermodynamic properties: the adiabatic exponent, the acoustic potential, the sound-speed gradient and the plasma coupling parameter. Section 3 presents the seismic analysis of the outer layers for all the six models and uncovers the impact that electrostatic interactions can produce on the acoustic spectrum. Finally, we summarise our conclusions in Section 4.

\begin{figure*} 
	\centering
	\includegraphics[width=0.30\textwidth]{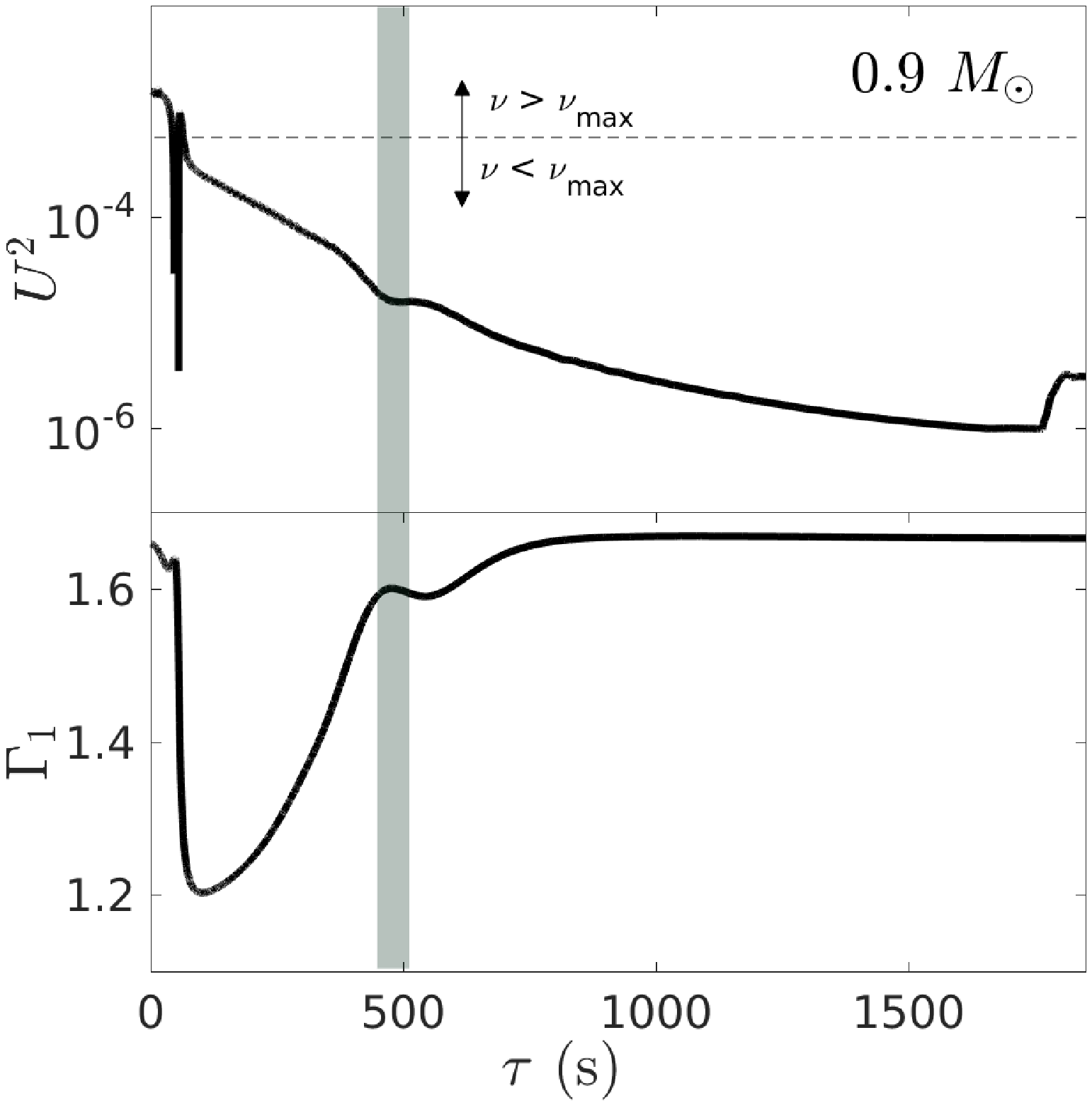}
	\includegraphics[width=0.30\textwidth]{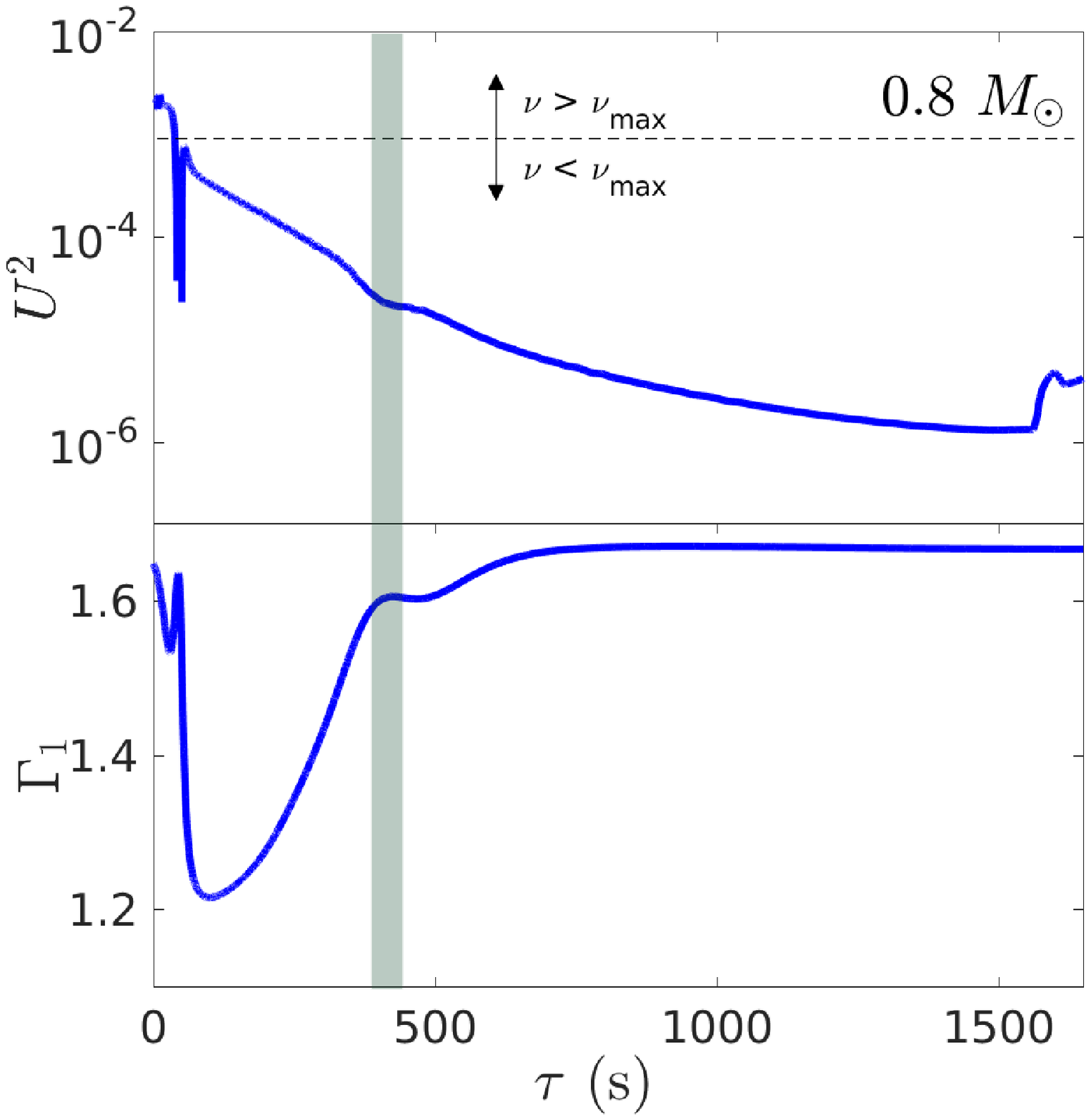}
	\includegraphics[width=0.30\textwidth]{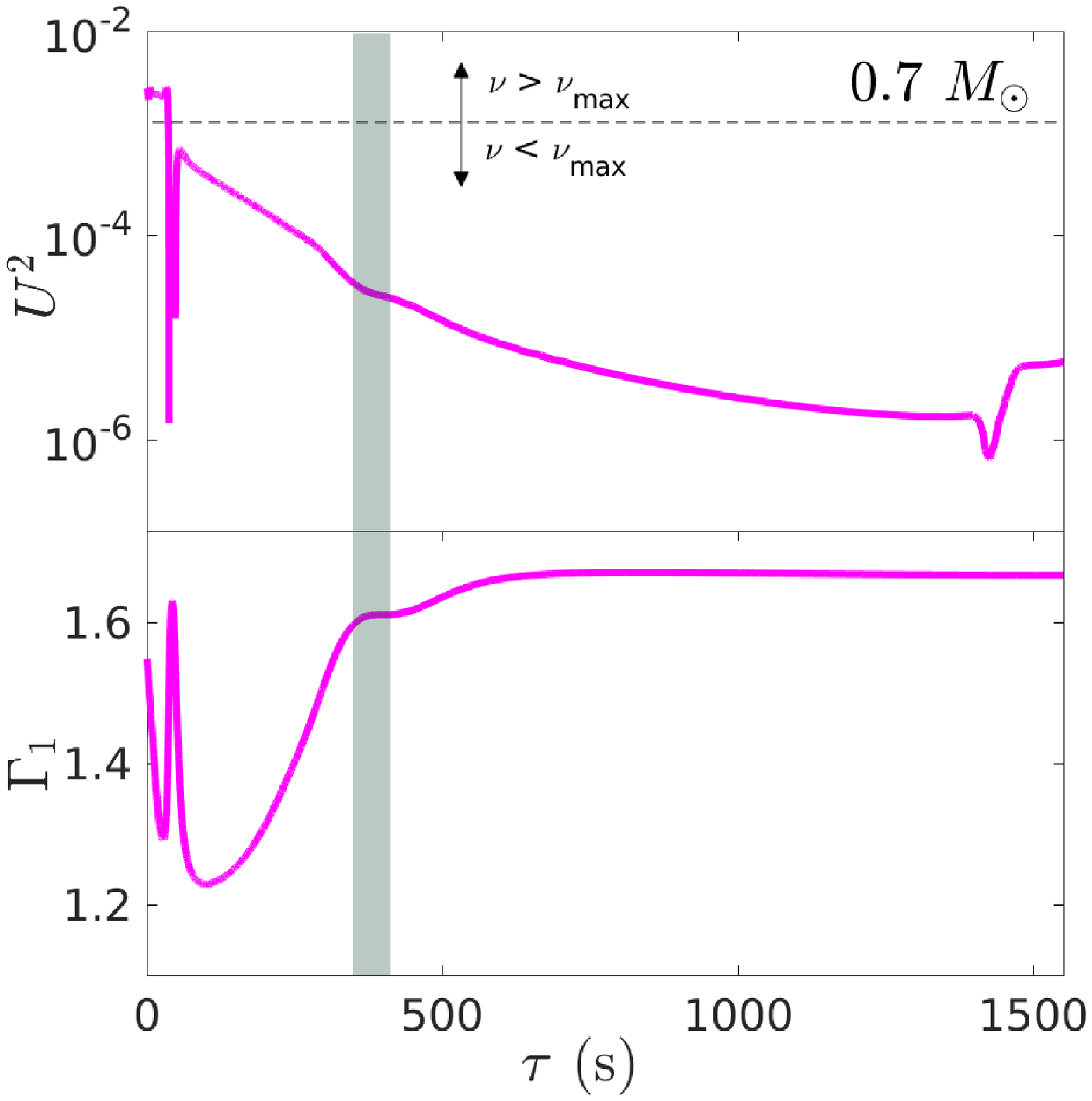}\\
	\includegraphics[width=0.30\textwidth]{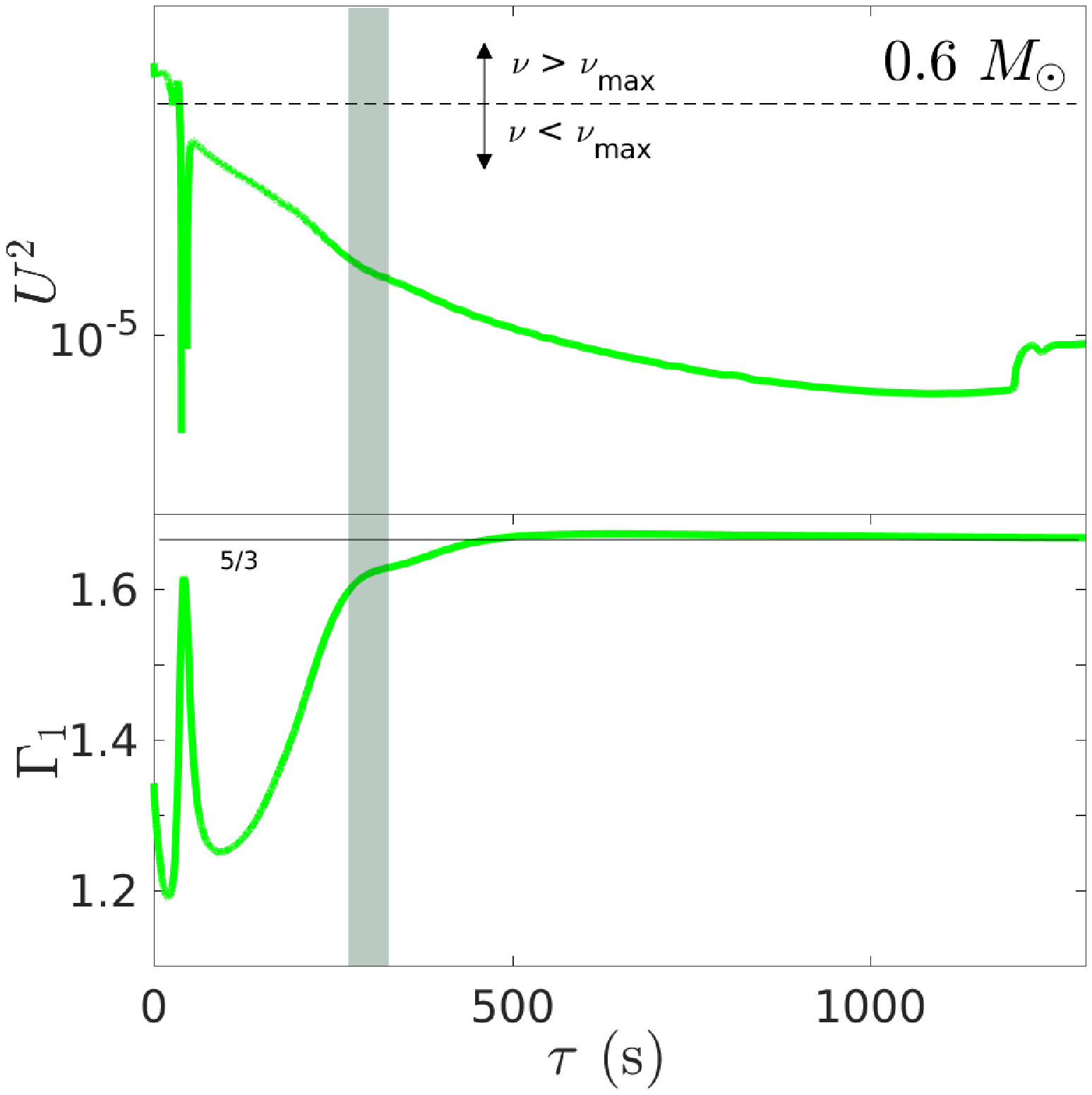}
	\includegraphics[width=0.30\textwidth]{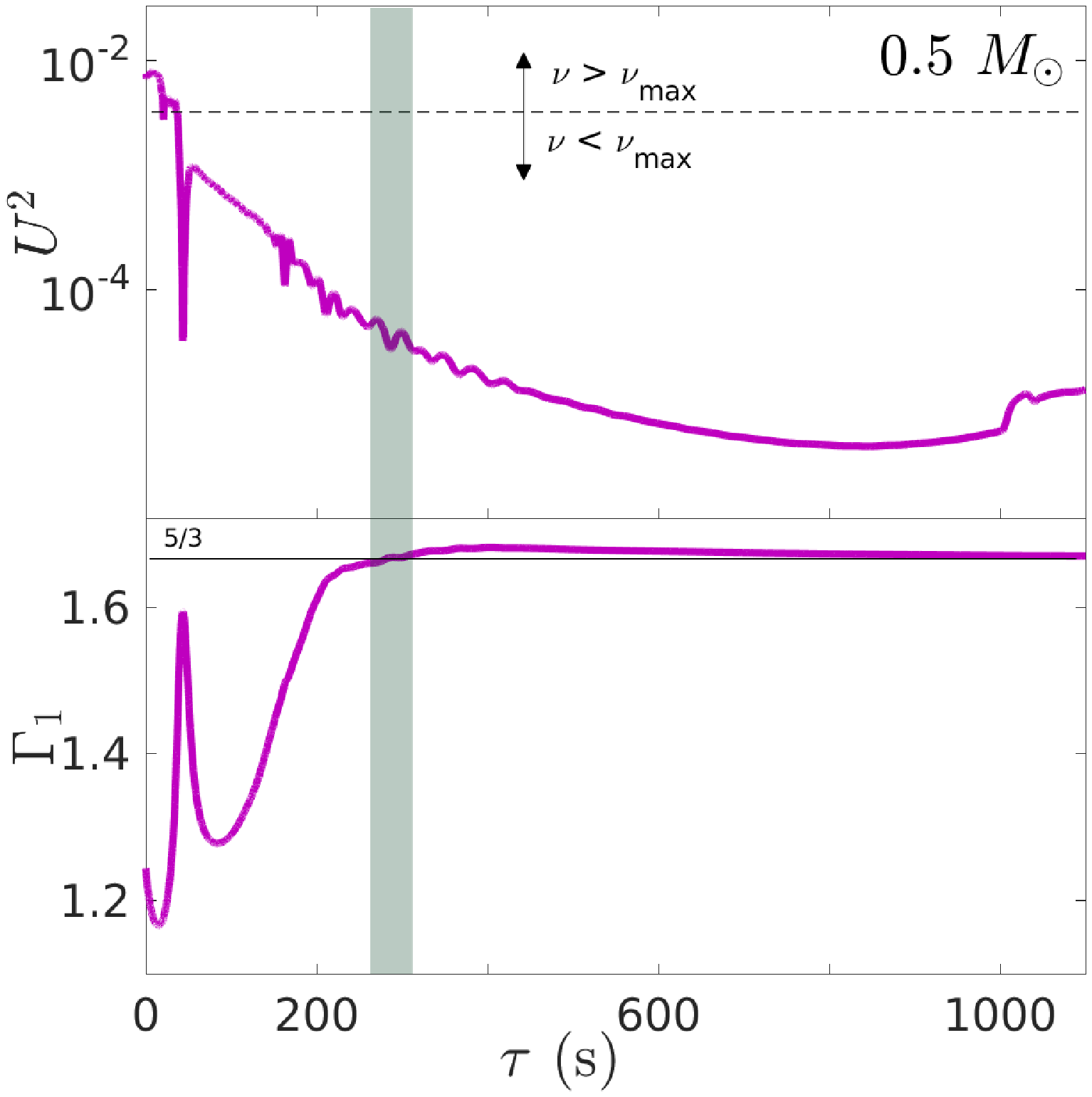}
	\includegraphics[width=0.30\textwidth]{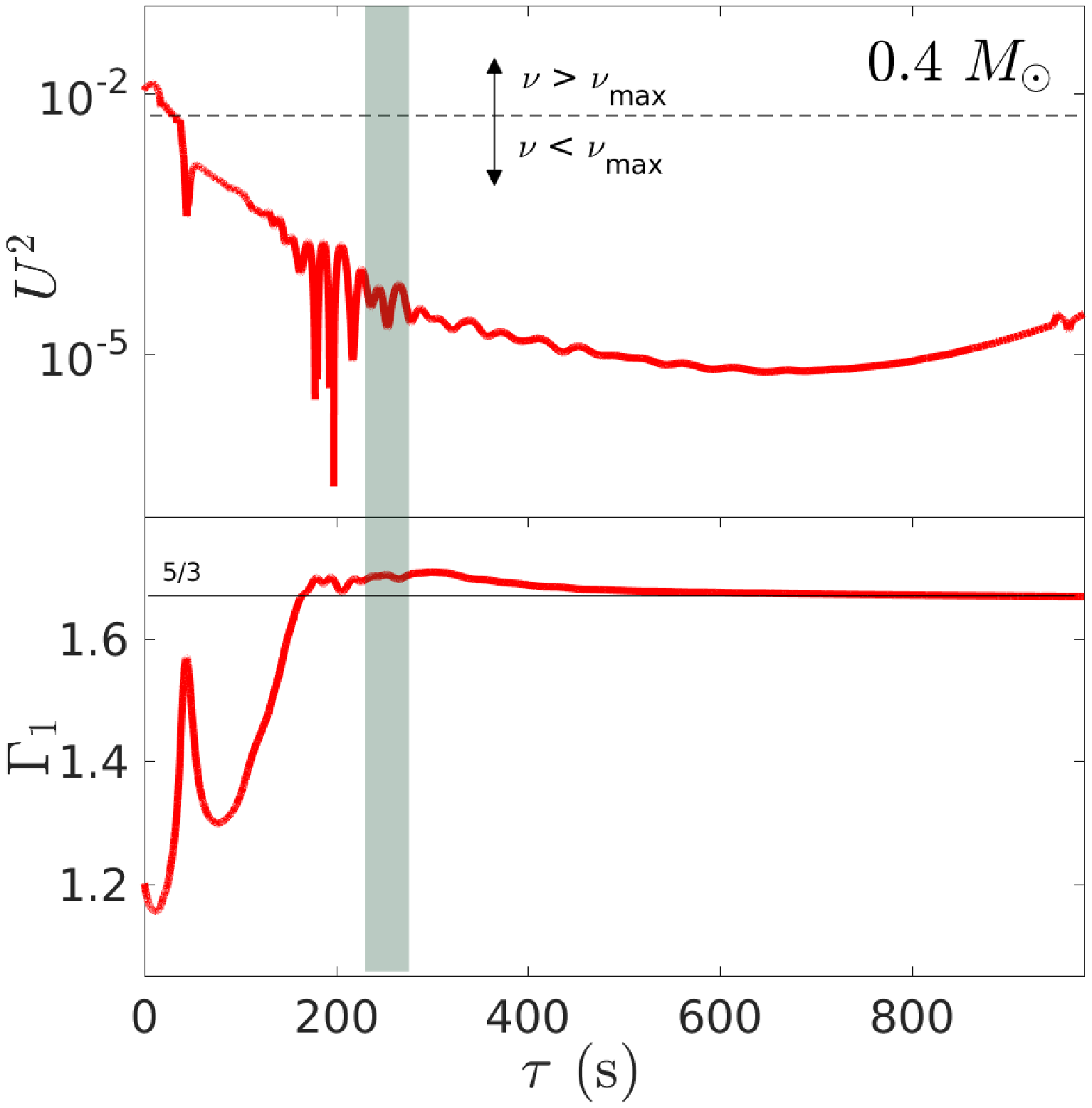}
	\caption{Representation of the acoustic potentials (top panels) and $\Gamma_1$ (bottom panels) for the outer convective layers of all the theoretical models. The vertical shadowed grey bar highlights the region where the ionisation fraction of single ionised helium reaches its maximum value. This is also the region where the peak/dip in $\Gamma_1$/$U^2$ is more pronounced for more massive models $(0.9 , \, 0.8,  \, \text{and} \; 0.7 \, M_\odot)$. Horizontal dashed lines represent the value of angular frequencies $\omega^2_{\text{max}} = (2 \pi \nu_{\text{max}})^2$ for each one of the models.}
	\label{fig2}
\end{figure*}

\section{The adiabatically stratified convective zones of low-mass stars}\label{sec2}
\subsection{The {\normalfont{MESA}} models}\label{subsec2.1}

This work is based on the results obtained from six main-sequence theoretical models of low-mass stars spanning from $0.9$ to $ 0.4 \, M_{\odot}$. These models were built with the one-dimensional (1D) stellar evolution code MESA v12115 \citep[Modules for Experiments in Stellar Astrophysics;][]{2019ApJS..243...10P, 2018ApJS..234...34P, 2015ApJS..220...15P, 2013ApJS..208....4P, 2011ApJS..192....3P}.

As it is expected from the theory of stellar structure and evolution all the main-sequence final models have a radiative core and a convective envelope, with the lower mass models having thicker convective zones as predicted \citep[e.g.][]{2004cgps.book.....W}. The details and choices for the input physics are as follows. All the six main-sequence models were evolved with solar metallicity adopting the chemical mixture of \citet{2009ARA&A..47..481A}. We used the MESA EOS tables which are based on the OPAL EOS tables of \citet{2002ApJ...576.1064R}. At lower temperatures, MESA EOS tables provide a smooth transition to the SCVH tables \citep{1995ApJS...99..713S}. Concerning opacities, the code was used with OPAL tables at high temperatures \citep{1996ApJ...464..943I} supplemented with low temperature opacities of \citet{2005ApJ...623..585F}. The nuclear reaction rates adopted are from the Joint Institute for Nuclear Astrophysics Reaction Library \citep[JINA REACLIB;][]{2010ApJS..189..240C}. 

Atomic element diffusion and gravitational settling are included in all models according to the prescription of \citet{1994ApJ...421..828T}. Convection was treated in the context of the mixing length theory as described by \citet{1958ZA.....46..108B} without overshoot. We used a Grey--Eddington formalism to link the outermost atmosphere layers to the model structure. The boundary condition was set deep in the atmosphere at a $\tau=100$, as this option leads to more realistic models of low-mass stars \citep{2016ApJ...823..102C}. Finally, the adiabatic oscillation mode frequencies of the low degree modes ($l=0,1,2$) were calculated for each stellar model structure using the stellar pulsation code GYRE v5.2. {The main properties of these models are summarized in table \ref{table1}.

\begin{table}
	\centering
	\caption{\bf Resulting model parameters: radii, effective temperatures and frequencies of maximum power.}
	\label{tab:example_table}
	\begin{tabular}{cccc} 
		\hline
		$M/M_\odot$ & $R/R_\odot$ & $T_{\text{eff}}$ (K) & $\nu_{\text{max}}$ (mHz)\\
		\hline
		0.9 & 0.85 & 5632 & 3.92\\
		0.8 & 0.74 & 5142 & 4.79\\
		0.7 & 0.66 & 4618 & 5.66\\
		0.6 & 0.58 & 4294 & 7.28\\
		0.5 & 0.44 & 4119 & 9.51\\
		0.4 & 0.36 & 4002 & 11.6\\
		\hline
	\end{tabular}
	\label{table1}
\end{table}

\subsection{Thermodynamic properties of the turbulent outer regions}\label{subsec2.2}

The interiors of low-mass stars are composed essentially of two kinds of charged particles: electrons and ions. Neutral atoms and molecular compounds can be found only in the very outermost layers of low-mass stars \citep[e.g.][]{2000ARA&A..38..337C}. In these outermost layers the temperature gradient 
\begin{equation} \label{eq1}
	\nabla = \frac{d \ln T}{d \ln P}
\end{equation}
operates in the superadiabatic regime. Here $T$ and $P$ represent temperature and pressure, respectively.  This is a region of utmost complexity where large uncertainties coexist related to different physical processes, such as turbulent convection, magnetic fields, surface opacities and excitation of acoustic oscillations \citep[e.g.][]{1997ApJ...474..790D, 2013ApJ...767...78T}. Excepting these outermost layers, the convective envelopes of low-mass stars follow a quasi-adiabatic stratification, i.e. its structure is mainly regulated by the thermodynamics of the stellar plasma, in particular by the temperature gradient (equation \ref{eq1}) that in deeper regions of the convective envelopes can be considered almost adiabatic, $\nabla \approx \nabla_{ad}$ \citep[e.g.][]{2004cgps.book.....W}, where
\begin{equation}\label{eq2}
	\nabla_{\text{ad}} = \left( \frac{\partial \ln T}{\partial \ln P} \right)_{\text{ad}}.
\end{equation}
Three main factors are known to shape the structure of the convective envelopes, and thus,  the structure of partial ionisation zones \citep[e.g.][]{1992MNRAS.257...32V}. They are the specific entropy, which depends on the value of the mixing length parameter $\alpha=l/H_p$, where $l$ is the mean mixing length and $H_p$ the pressure scale height; the chemical composition; and finally, the equation of state (EOS). In this study, we assumed the same value of the mixing length parameter ($\alpha=1.8$) for all the models and the same chemical composition (as described in Section \ref{subsec2.1}). 

\begin{figure*}
	\centering
	\includegraphics[width=0.7\textwidth]{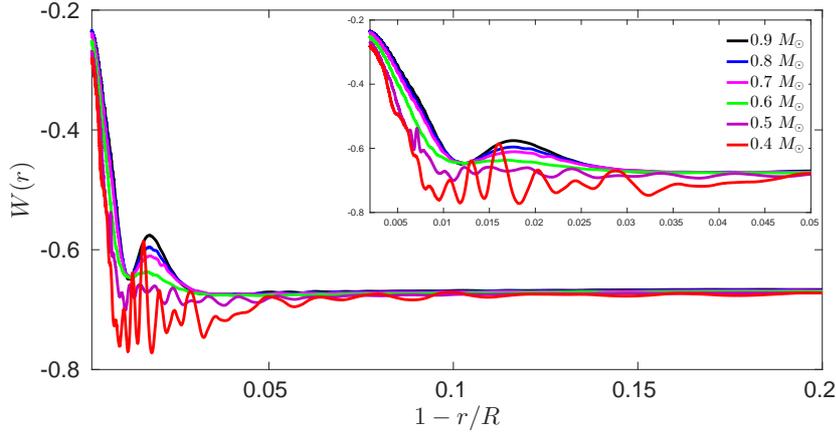}
	\caption{The sound speed gradient, $W(r)$, for the outer region of all models (up to 20 per cent under the surface of the star). The inset represents the first 5 per cent of the outer region. This is a particularly important region for partial ionisation processes where helium looses its two electrons. This is also the region where electrostatic interactions impact more the sound speed of the less massive stars.}
	\label{fig3}
\end{figure*}

Hence, the first adiabatic exponent, $\Gamma_1$, given by the relation
\begin{equation}\label{eq3}
	\Gamma_1= \left( \frac{\partial \ln P}{\partial \ln \rho} \right)_{\text{ad}}
\end{equation}
is entirely determined by the EOS and defines the structure of the adiabatic convective zones of the theoretical models. The behaviour of $\Gamma_1$ around its monatomic ideal gas value, $5/3$,  is influenced by two fundamental aspects \citep[e.g.][]{2007AIPC..948..213B}. First, the increase of the degree of ionisation decreases the $\Gamma_1$ value, causing a known depression in the radial profile of $\Gamma_1$, particularly throughout the partial ionisation zones of light elements, hydrogen and helium \citep[e.g.][]{2005MNRAS.361.1187M}. The second aspect that influences the behaviour of $\Gamma_1$ relates to the electrostatic interactions between charged particles, known generally as Coulomb effects. Coulomb effects increase the value of $\Gamma_1$ causing it to become larger than $5/3$ in some regions of the convective zone. Briefly, this can be explained as follows. Electrostatic interactions between particles reduce the pressure because charged particles tend to surround themselves by particles of opposite charge, yielding a decrease in the value of repulsive forces. This leads to a negative contribution of pressure to the EOS \citep[e.g.][]{2000A&A...364..157B, 2000MNRAS.316...71B}.

\begin{figure*} 
	\centering
	\includegraphics[width=0.30\textwidth]{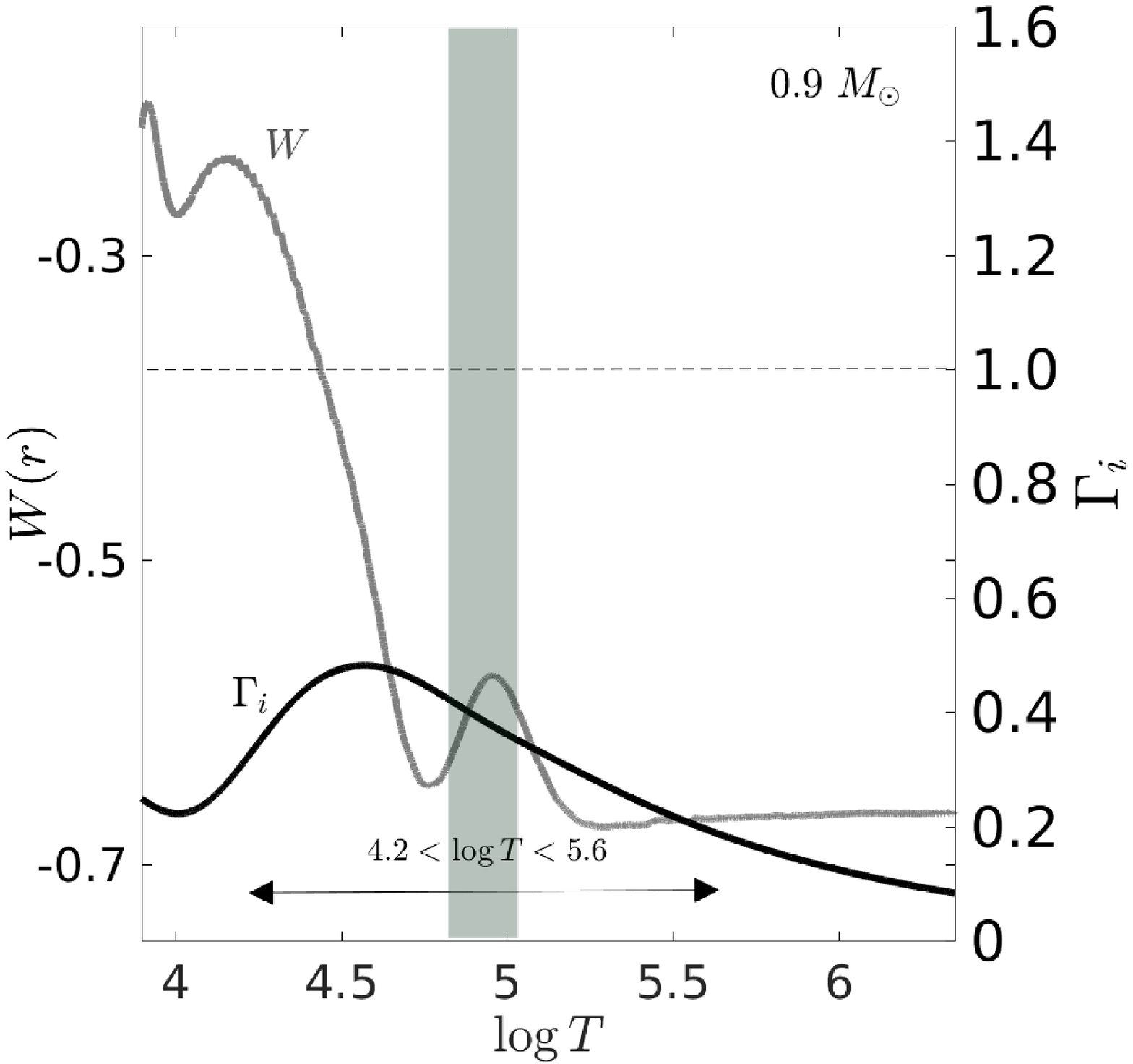}
	\includegraphics[width=0.30\textwidth]{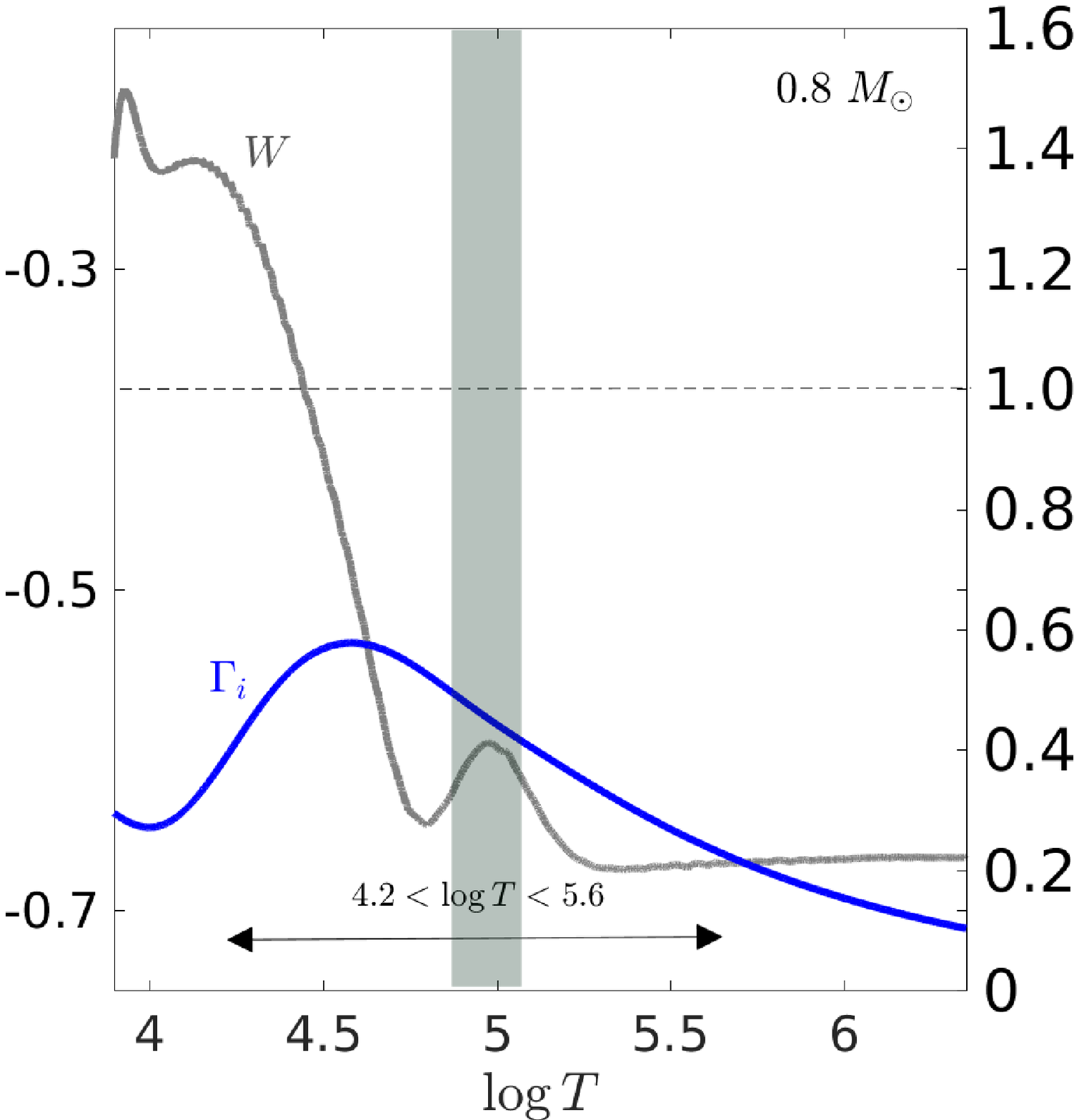}
	\includegraphics[width=0.30\textwidth]{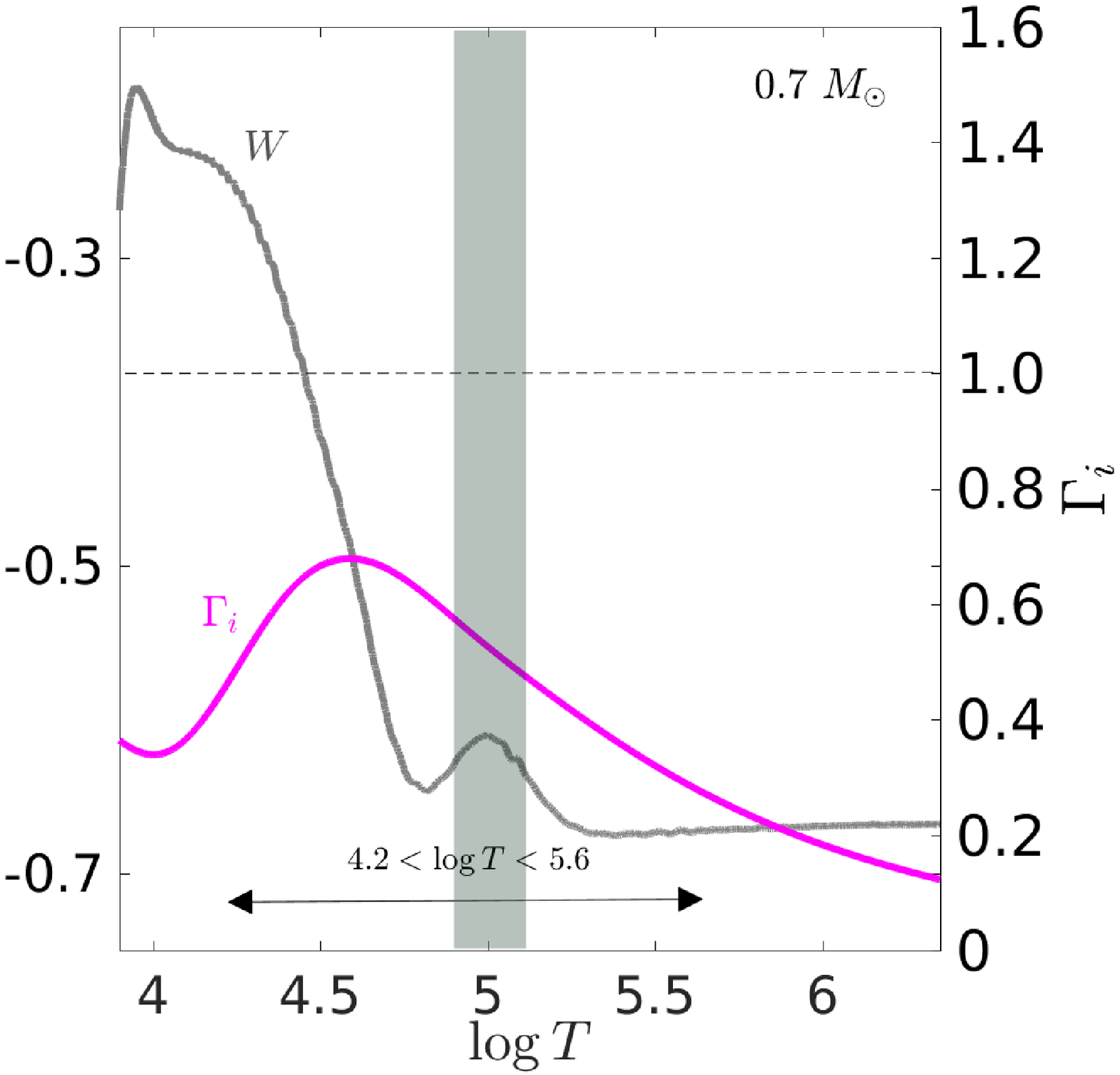}\\
	\includegraphics[width=0.30\textwidth]{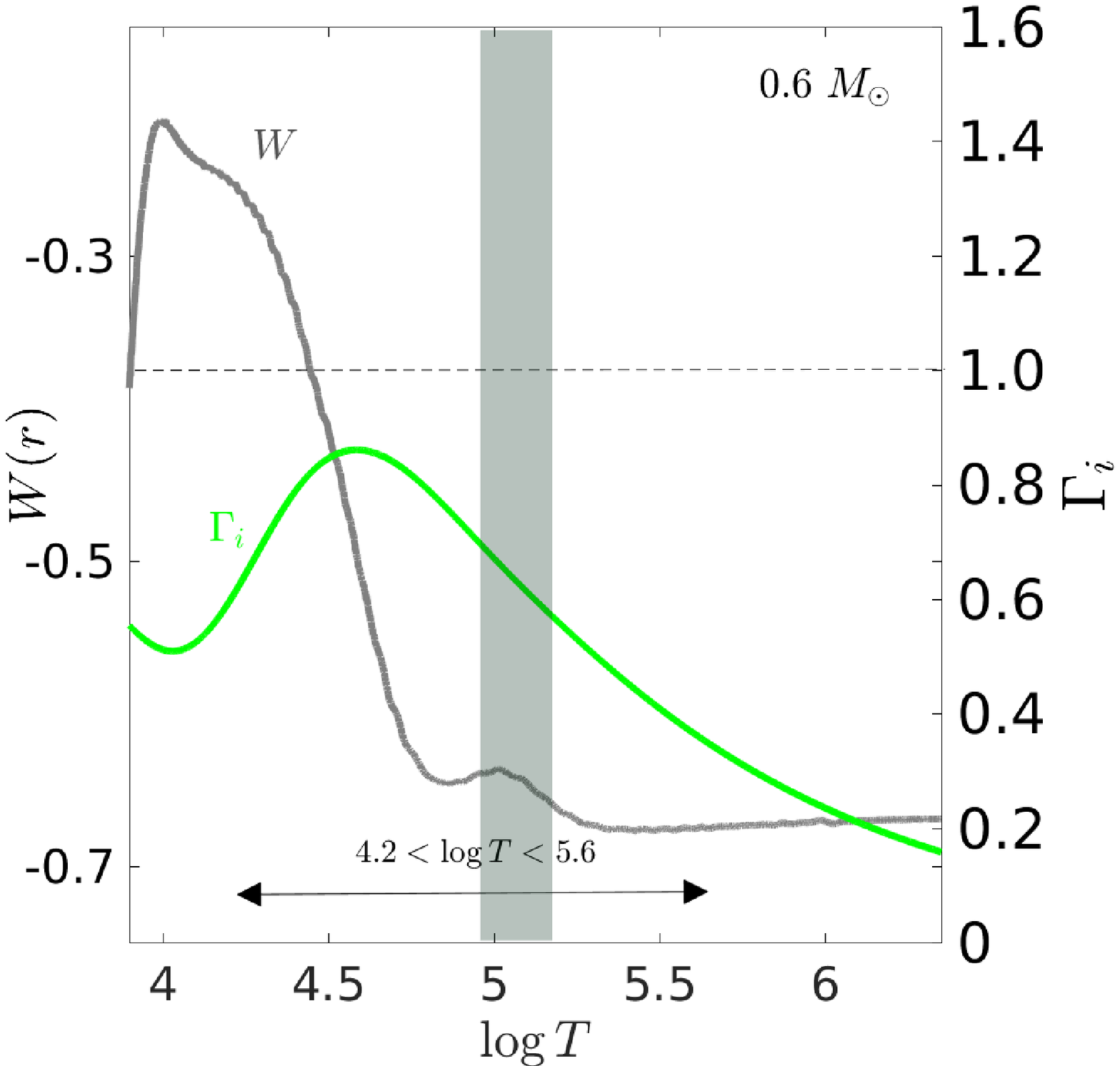}
	\includegraphics[width=0.30\textwidth]{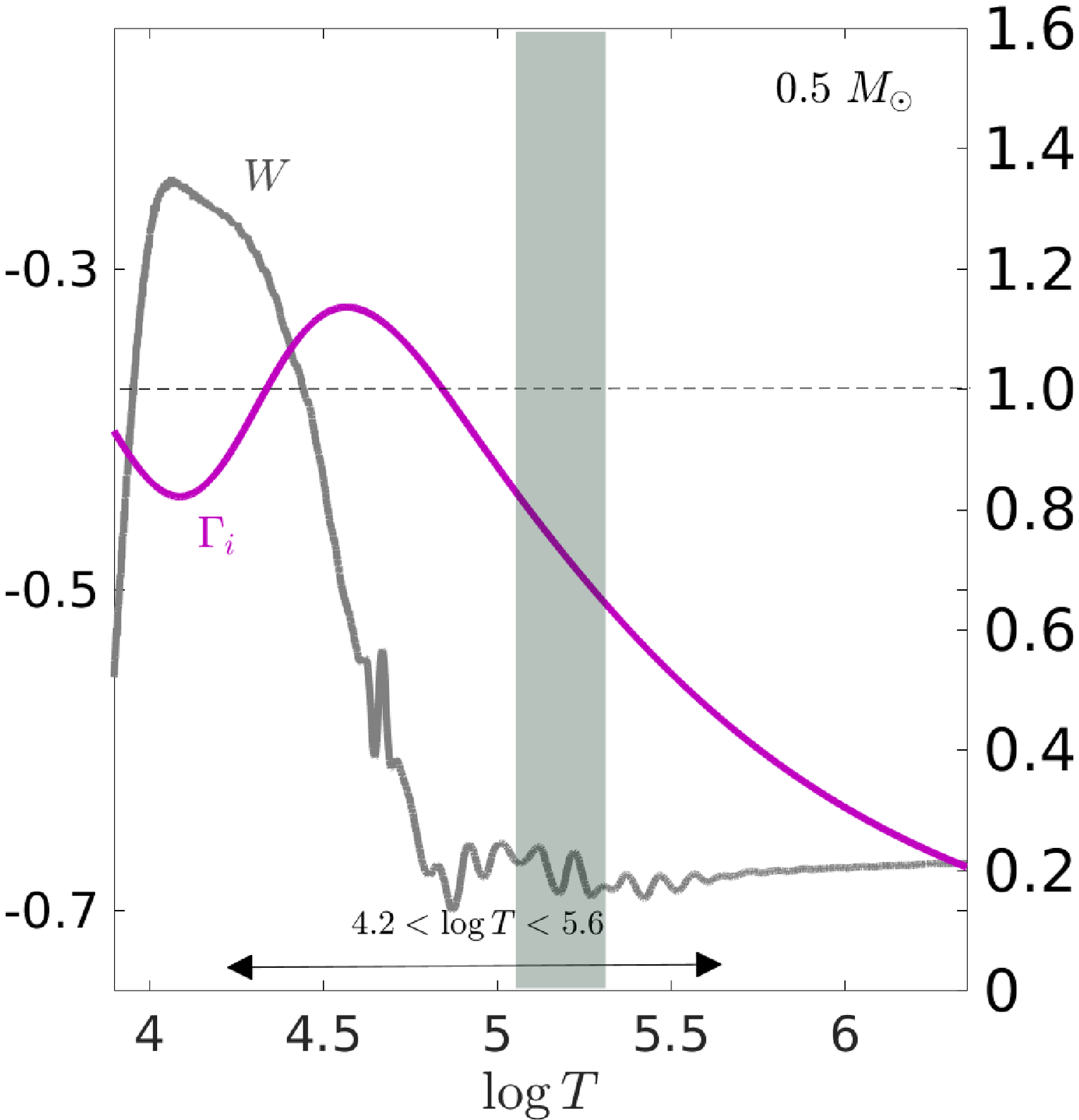}
	\includegraphics[width=0.30\textwidth]{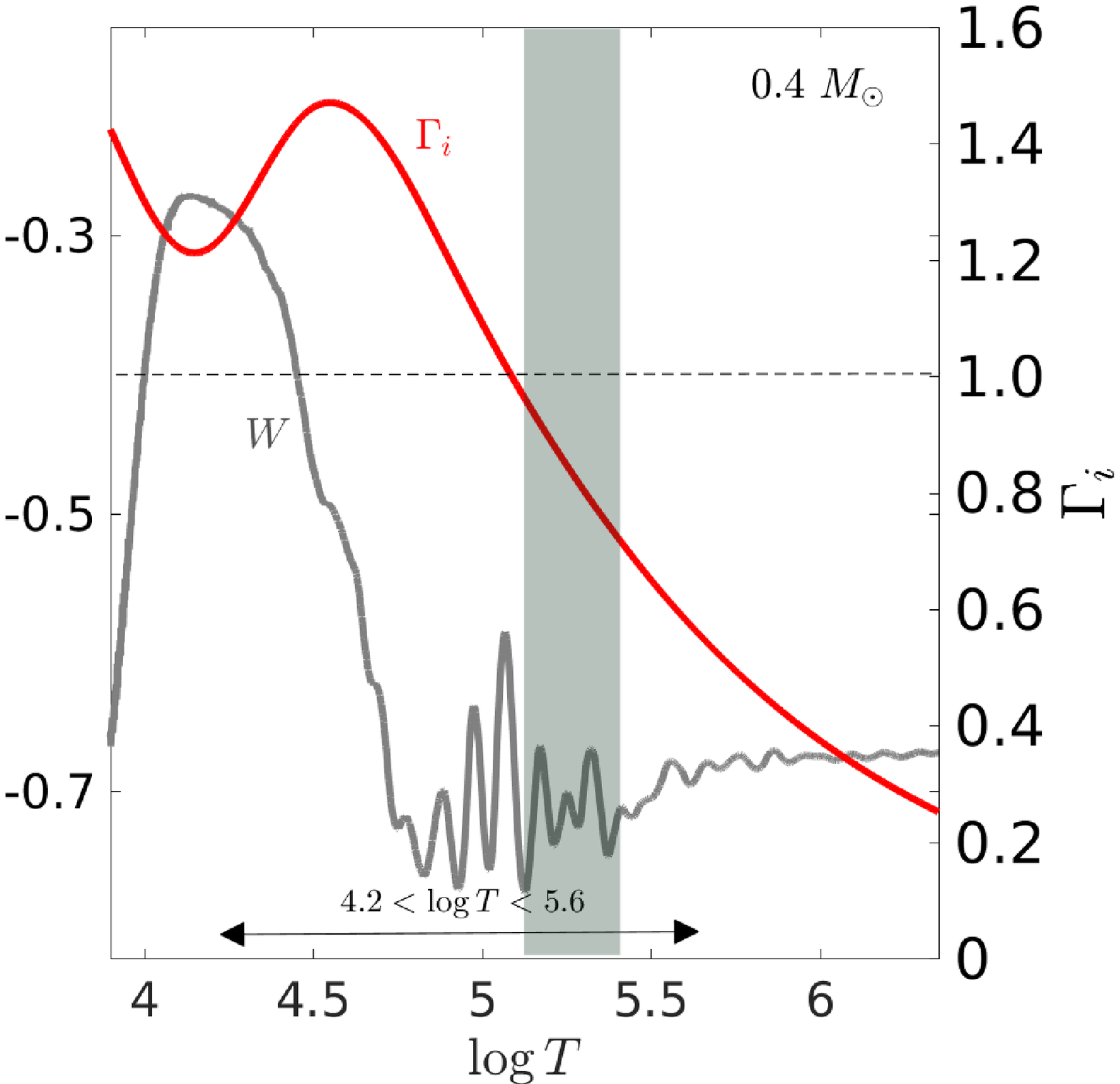}
	\caption{The ionic plasma interaction parameter, $\Gamma_i$, as a function of temperature ($\log T$). Also plotted in this figure, with grey colour, is the sound speed gradient, $W(r)$. The sharp features shown by lower-mass models correspond to the regions where the plasma interaction parameter is close or greater than one. The vertical grey bar signalises the region where single ionised helium reaches its maximum value.}
	\label{fig4}
\end{figure*}

Figure \ref{fig1} represents $\Gamma_1$ as a function of the fractional radius for the set of computed models. The well-known depression of $\Gamma_1$ associated with the helium partial ionisation is clearly visible for more massive stars $(0.9 , \, 0.8,  \, \text{and} \; 0.7 \, M_\odot)$. For the $0.6$ and $0.5 \, M_\odot$ stellar models, the $\Gamma_1$ depression almost vanishes. We found this effect to be related to the increase of density and growing influence of Coulomb effects. The lowest mass model, $0.4 \, M_\odot$, shows a complex pattern of interplay between the two main aspects that influence the first adiabatic exponent, partial ionisation, and Coulomb effects, with the latter revealing an exceptional impact for this model.

$\Gamma_1$ also plays a special role in asteroseismological studies. This is because, for acoustic oscillations, the adiabatic sound speed, $c$, of the sound wave is given by
\begin{equation}\label{eq4}
	c=\sqrt{\frac{\Gamma_1 P}{\rho}} \, ,
\end{equation}
where $P$ and $\rho$ stand for pressure and density, respectively. The sound speed profile will be strongly dependent on the same processes that impact $\Gamma_1$. 
All stars with outer convective envelopes have the potential to excite solar-like oscillations. These are sound waves that are reflected by the surface of the stars and propagate throughout the stellar interiors up to different depths depending on the properties of the reflection \citep[e.g.][]{2013ARA&A..51..353C}. The quantity
\begin{equation}\label{eq5}
	\tau=\int_r^R \frac{dr}{c}
\end{equation}
is called the acoustic depth and represents the traveltime of the sound wave from the surface of the star to a depth of radius $r$. As shown by \citet{1989ASPRv...7....1V}, the propagation of such oscillations in the outer parts of convective zones can be described by a Schr\"{o}dinger-type equation
\begin{equation}\label{eq6}
	\frac{d^2 \psi}{d\tau^2} + (U-\omega^2)\psi = 0 \, ,
\end{equation}
where $\omega$ is the angular frequency of the acoustic mode, $\psi$ a wave function determined by the radial displacements, and $\tau$ the acoustic depth (given by equation \ref{eq5}) as an independent variable. Equation \ref{eq6}, which describes the adiabatic nonradial oscillations in the outer layers, is obtained within two main approximations: one is the Cowling approximation, possible because in surface layers the gravitational effects can be neglected; and the other approximation takes into account that in these regions and for low-degree modes, the radial component of the wavenumber is much greater than the horizontal component, and thus, the trajectories of the sound waves can be considered almost vertical near the surface and dependent on the frequency alone (i.e., $l=0$, with $l$ being the mode degree). The potential, $U^2$, usually known as the acoustic potential is defined as
\begin{dmath}\label{eq7}
	U^2 = N^2 - \frac{c}{2} \frac{d}{dr} \left[ c \left( \frac{2}{r} + \frac{N^2}{g} - \frac{g}{c^2} - \frac{1}{2c^2} \frac{dc^2}{dr} \right)  \right] + \frac{c^2}{4} \left( \frac{2}{r} + \frac{N^2}{g} - \frac{g}{c^2} - \frac{1}{2c^2} \frac{dc^2}{dr}   \right)^2 \, .
\end{dmath}
Here, $N$ is the buoyancy frequency, $g$ is the gravitational acceleration and $c$ the adiabatic sound speed defined by equation \ref{eq4}. In convective zones with adiabatic stratifications, the acoustic potential is determined by the model structure and can be written as an expression explicitly dependent on the first adiabatic exponent, $\Gamma_1$, as shown by \citet{1991SoPh..133..149M}. Therefore, physical processes like partial ionisation of atomic species and electrostatic interactions between the charged particles that constitute the stellar plasma will impact the acoustic potential of the theoretical models. 

Figure \ref{fig2} represents, for each model, the acoustic potential (top panel) and the first adiabatic exponent (bottom panel). For more massive stars $(0.9 , \, 0.8,  \, \text{and} \; 0.7 \, M_\odot)$
is possible to identify, in the acoustic potential, the hump associated with the region of the helium partial ionisation.
This hump was well studied in the solar case \citep[e.g.][]{1989SvAL...15...27B, 2001MNRAS.322..473L} and has recently been investigated also for solar-type stars \citep{2017ApJ...843...75B, 2018ApJ...853..183B}. Because density increases when the mass of the model decreases, the relevance of Coulomb effects also increases and the hump becomes less pronounced towards less massive stars. Hence, the acoustic potential of the $0.6 \, M_\odot$ model is almost unperturbed in the partial ionisation region of helium. For less massive models, $0.5$ or $0.4 \, M_\odot$, the acoustic potential reveals a pattern of perturbations with different characteristics. Whereas for stars with masses greater than $0.6 \, M_\odot$ the perturbed region in the acoustic potential is a broad and "smooth" hump, for stars with masses lesser than $0.6 \, M_\odot$, the perturbed region in the acoustic potential consists of a broad "noisy" region. This means that Coulomb effects in low-mass stars can strongly perturb the sound speed leading to new possibilities of diagnostics related to the uncertainties of the EOS.

\begin{figure*} 
	\centering
	\includegraphics[width=0.30\textwidth]{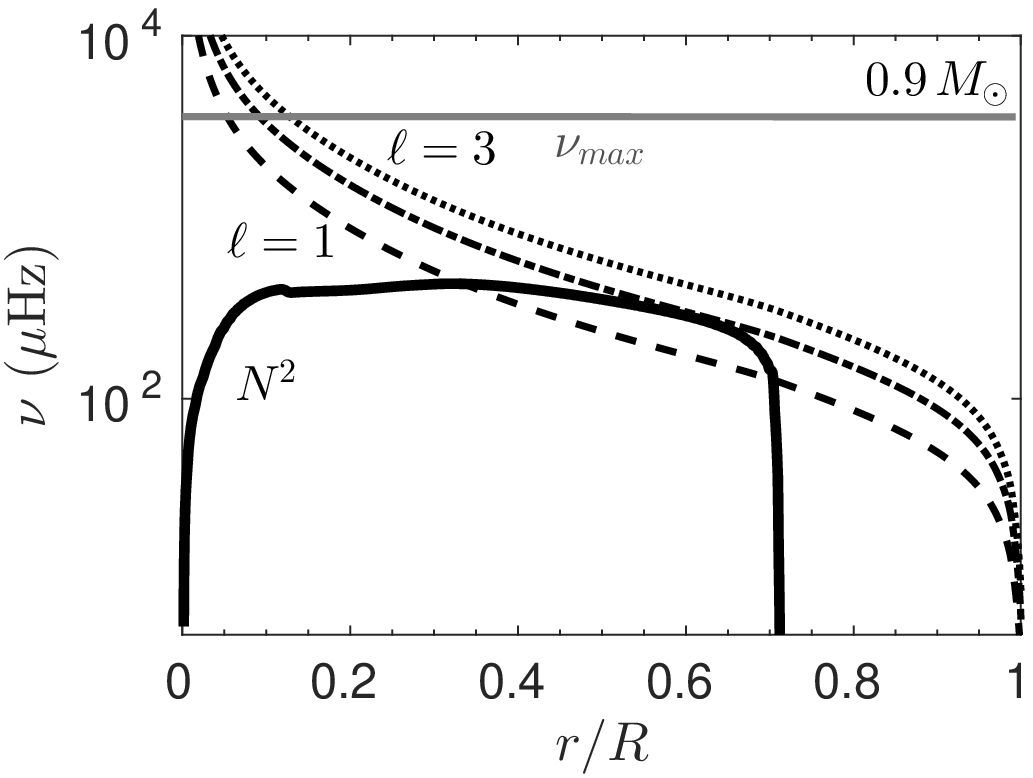}
	\includegraphics[width=0.30\textwidth]{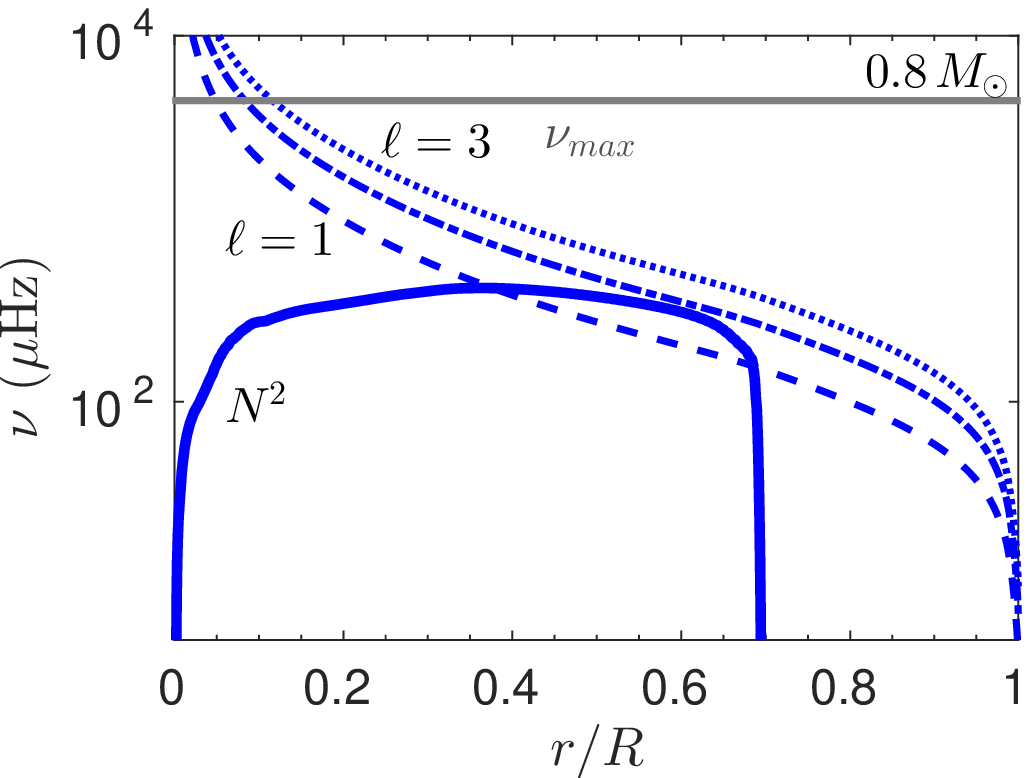}
	\includegraphics[width=0.30\textwidth]{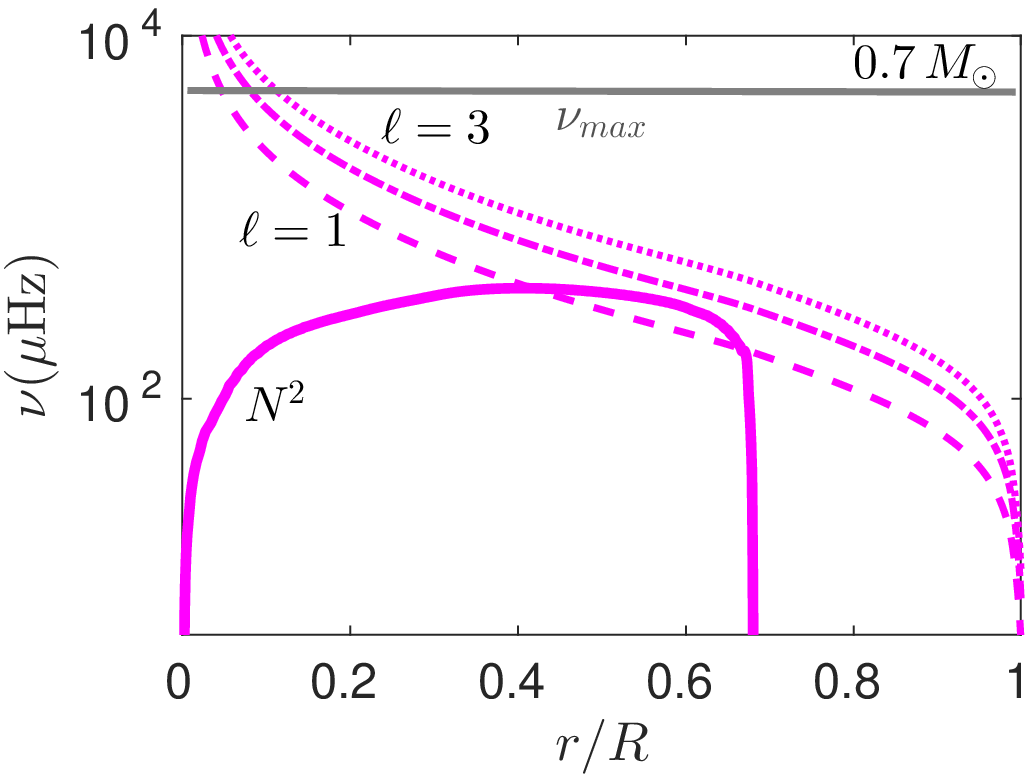}\\
	\includegraphics[width=0.30\textwidth]{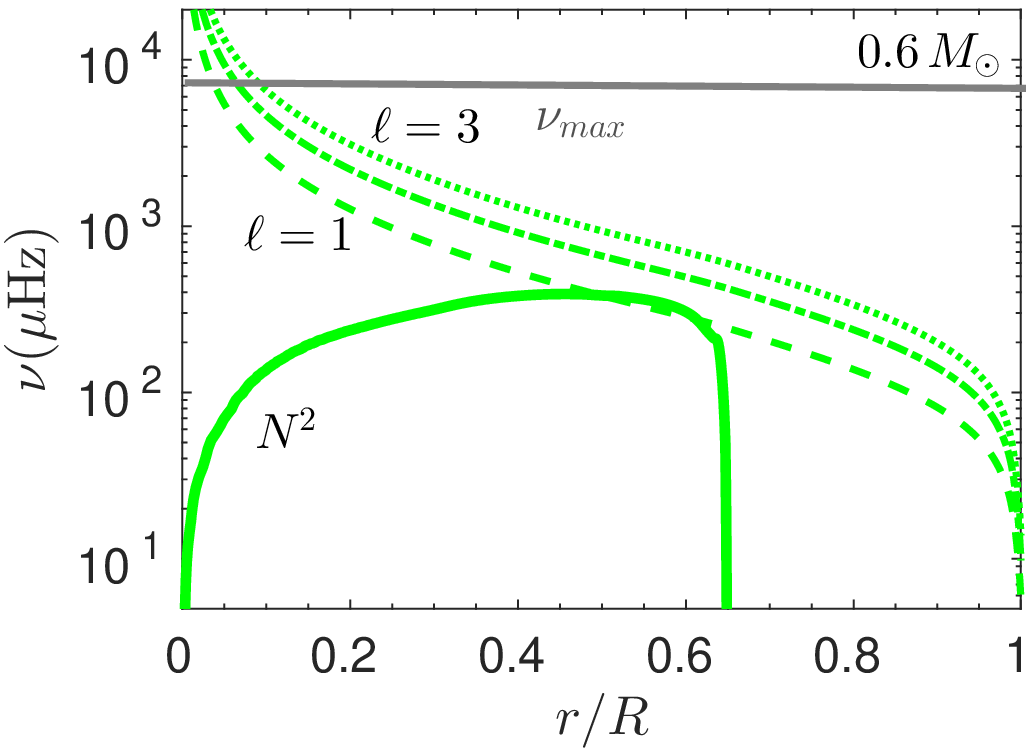}
	\includegraphics[width=0.30\textwidth]{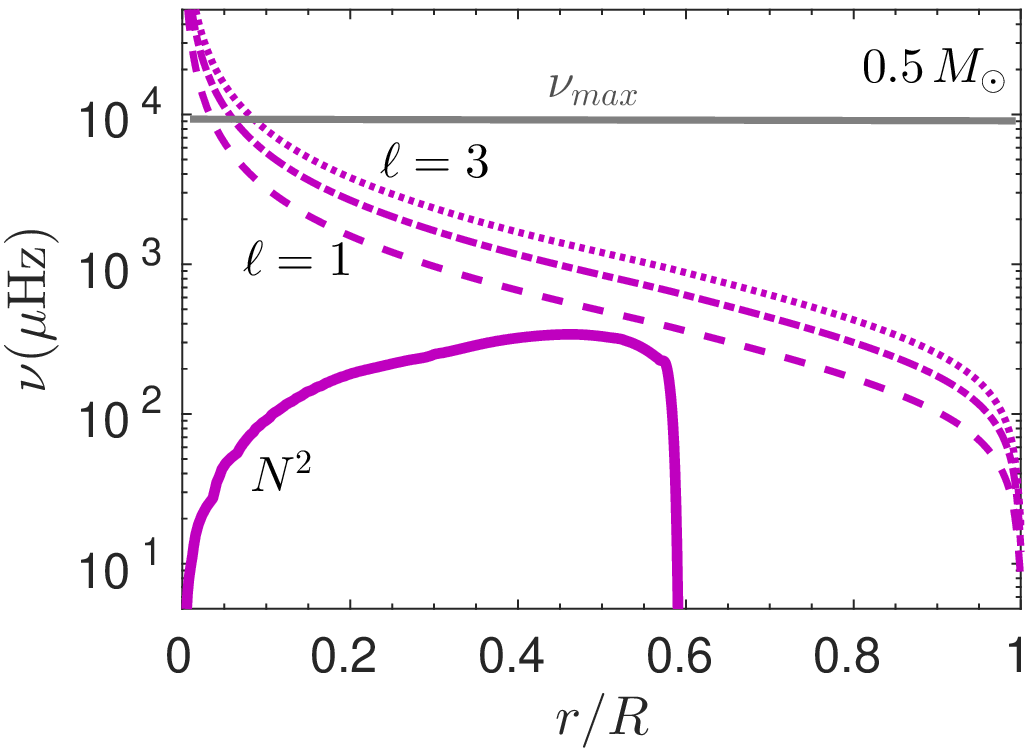}
	\includegraphics[width=0.30\textwidth]{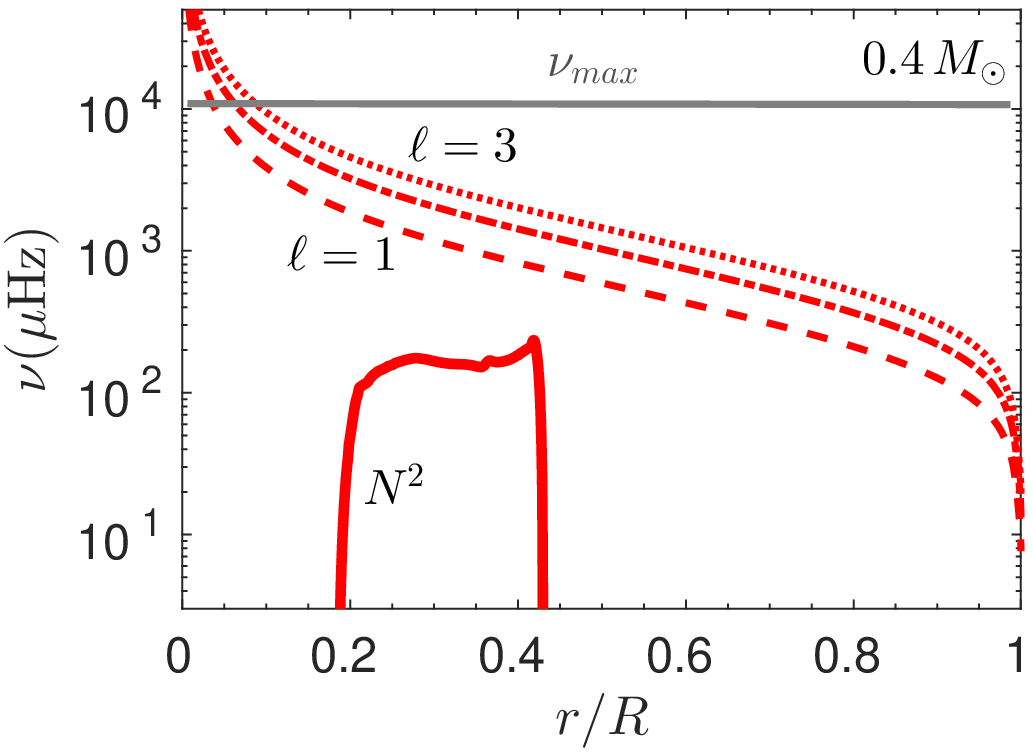}
	\caption{Propagation diagrams for all the main-sequence stellar models. The solid line is the Brunt--V\"{a}is\"{a}l\"{a}, while dashed ($l=1$), dot-dashed ($l=2$), and dotted ($l=3$) lines show the Lamb frequency. A region will be a propagation zone if $\nu > S_l/{(2\pi)}$ and $\nu > N^2/{(2\pi)}$ at the same time, or $\nu < S_l/{(2\pi)}$ and $\nu < N^2/{(2\pi)}$ simultaneously. The horizontal grey line indicates the frequency of maximum power, $\nu_{\text{max}}$, obtained from scaling relations in each case.}
	\label{fig5}
\end{figure*}

The diagnostic quantity,
\begin{equation}\label{eq8}
	W(r) = \frac{r^2}{Gm(r)} \frac{dc^2}{dr} \, ,
\end{equation}
known as the dimensionless sound speed gradient \citep[e.g.][]{2015SSRv..196...49B}, is a diagnostic, particularly sensitive to changes in $\Gamma_1$ exponent as shown by \citet{1984MmSAI..55...13G}. Being sensitive to $\Gamma_1$, this quantity reveals details of the changes occurring in ionisation zones \citep[e.g.][]{2000MNRAS.316...71B} and also, in this particular case of low-mass models, changes related to non-ideal Coulomb effects. 
Figure \ref{fig3} illustrates the paradigm discussed above concerning the two relevant physical processes that can impact the acoustic oscillations. For more massive stars, the perturbations of the acoustic spectrum will be more influenced by partial ionisation processes, leading to a moderate variation in the sound speed gradient. Instead, for less massive stars, the acoustic spectrum will be more influenced by Coulomb effects, leading to a strong oscillatory component in the profile of the sound speed gradient. 

It is well known that even for solar models, Coulomb interactions between particles in the gas should be taken into account in the EOS, although their effects are not so significant as the effects of partial ionisation in the convective zone \citep[e.g.][]{2020arXiv200706488C}. For low-mass stars, Coulomb interactions become relevant and dominant even in partial ionisation regions as we can see from figure \ref{fig1}. A measure of how important these effects are in the interior of a stars can be expressed by the plasma coupling parameter, $\Gamma_i$ given as the ratio of the average potential energy to the thermal kinetic energy in the gas,
\begin{equation}\label{eq9}
	\Gamma_i = \frac{(Z e)^2}{a_i k_B T}
\end{equation}
where
\begin{equation}\label{eq10}
	a_i = \left( \frac{4}{3} \pi n_i    \right)^{-\frac{1}{3}}
\end{equation}
is the Wiegner-Seitz radius, which represents the mean inter-ion distance. Here, $e$ is the electron charge, $k_B$ the Boltzmann constant, $T$ the temperature, $Z$ the mean ionic charge, and $n_i$ the ionic density. This is the well-known ionic coupling parameter \citep[e.g.][]{1977PhRvA..16.2153H, 1986ApJS...61..177P, 2009pfer.book.....M}, which gives us the relative importance of the interactions between ions. Figure \ref{fig4} shows the plasma interaction parameter for all the models where it is most relevant: the outer layers, namely in the temperature range $4.2 \lesssim \log T \lesssim 5.6$. In this temperature range, the electrostatic interactions are known to be more significant, as this is the interval of temperatures where the two helium ionisations occur \citep[e.g.][]{1991LNP...388...11C, 1992A&ARv...4..267C, 1999ApJ...518..985B}. As expected for less massive models, Coulomb effects are extremely important in the partial ionisation regions of light elements. In some regions, $\Gamma_i$ exceeds the unity threshold value where the thermal and electrostatic energies have the same importance. The sound speed gradient given by equation \ref{eq8} is also shown in Figure \ref{fig4}. The sharp features of the lowest mass models corresponding to a strong scatter of the acoustic modes can be easily related to the highest values (values close or greater than 1) of the plasma interaction parameter. Therefore, the local properties of this scattering reveal the underlying physics of nonideal effects. The next section will cover the impact of electrostatic interactions on the acoustic oscillations in the more external parts of the stellar models.

\section{Seismology of low-mass stars: a study for the outer convective regions} \label{sec3}

\begin{figure*} 
	\centering
	\includegraphics[width=0.30\textwidth]{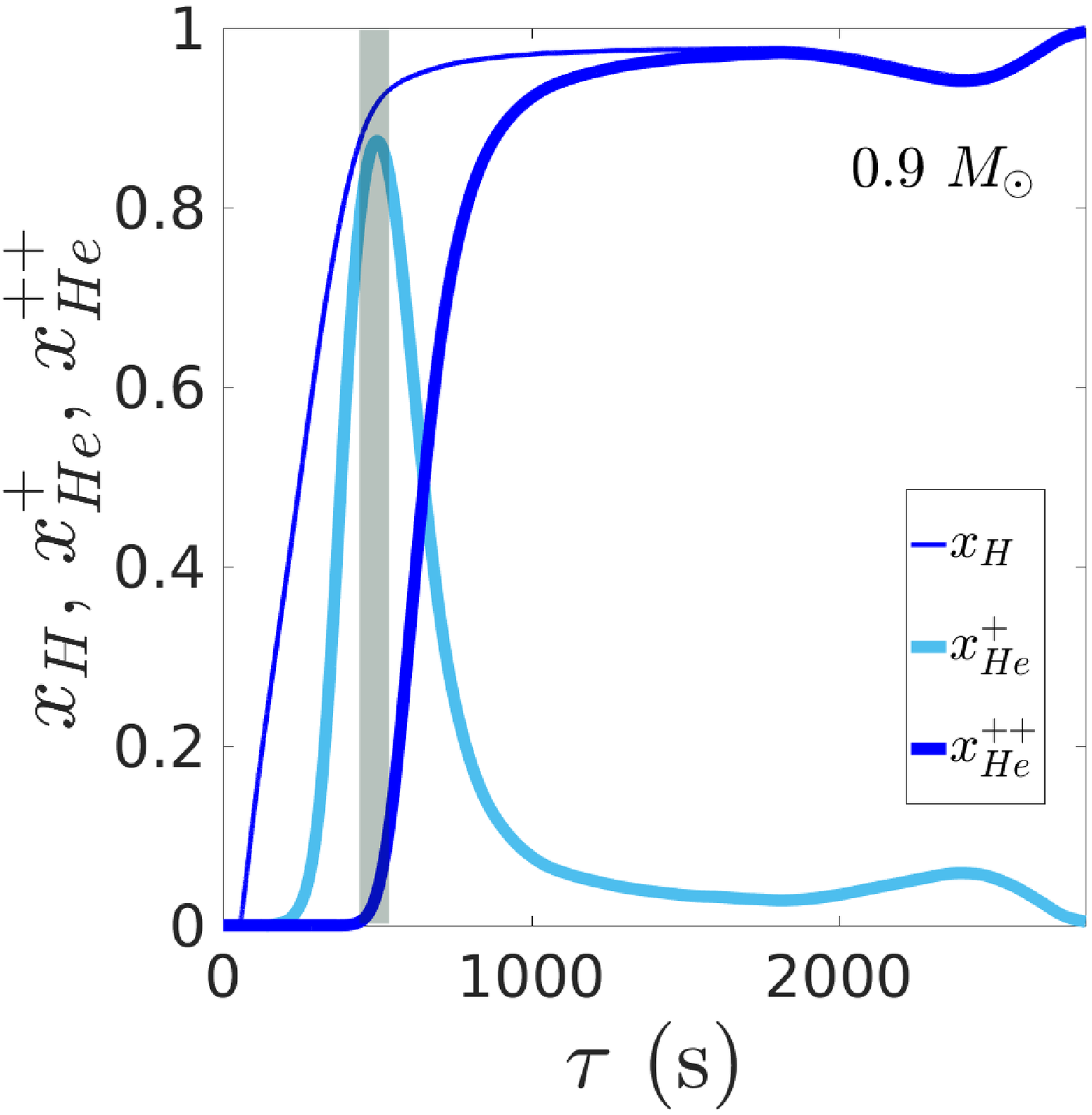}
	\includegraphics[width=0.30\textwidth]{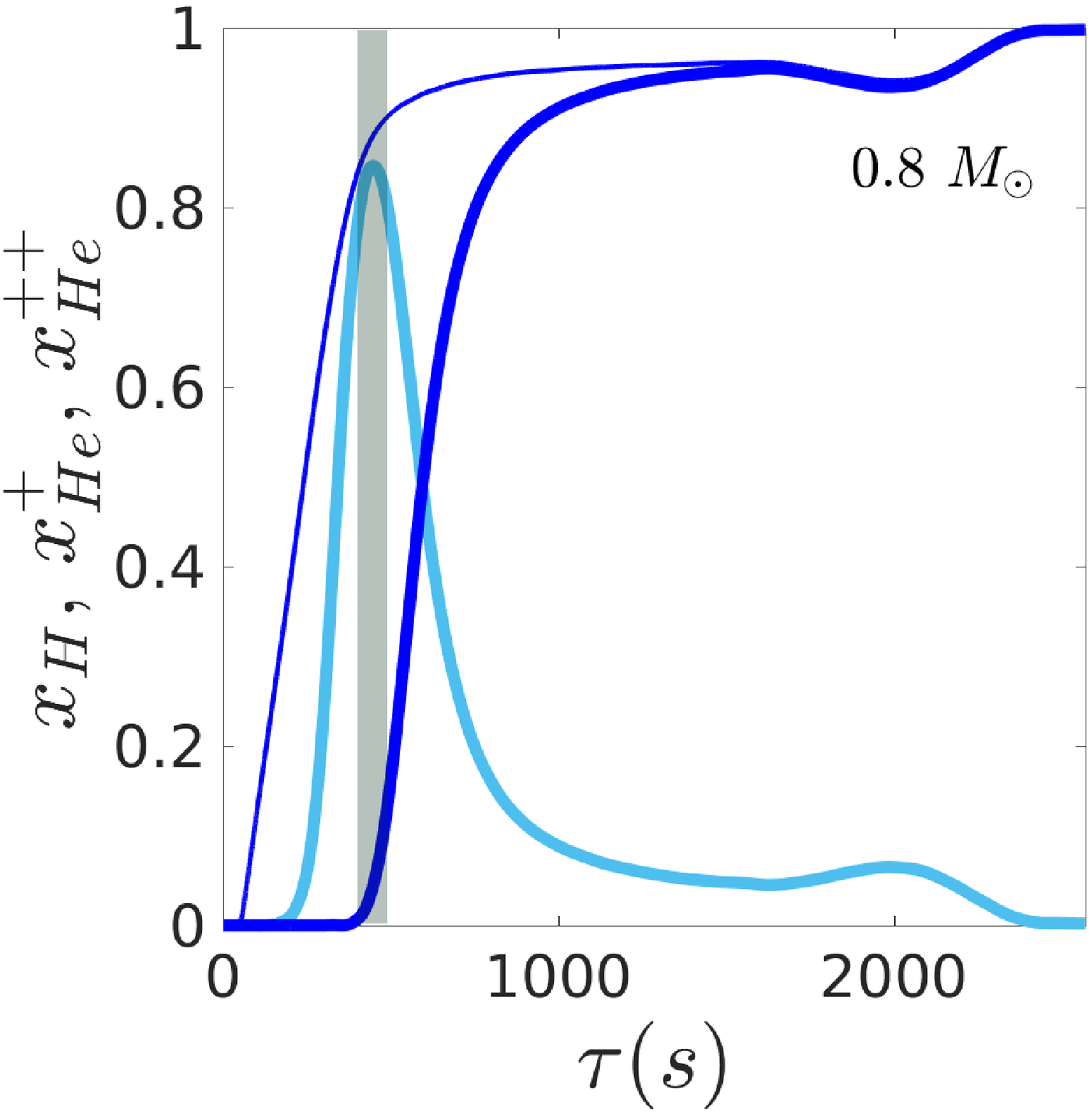}
	\includegraphics[width=0.30\textwidth]{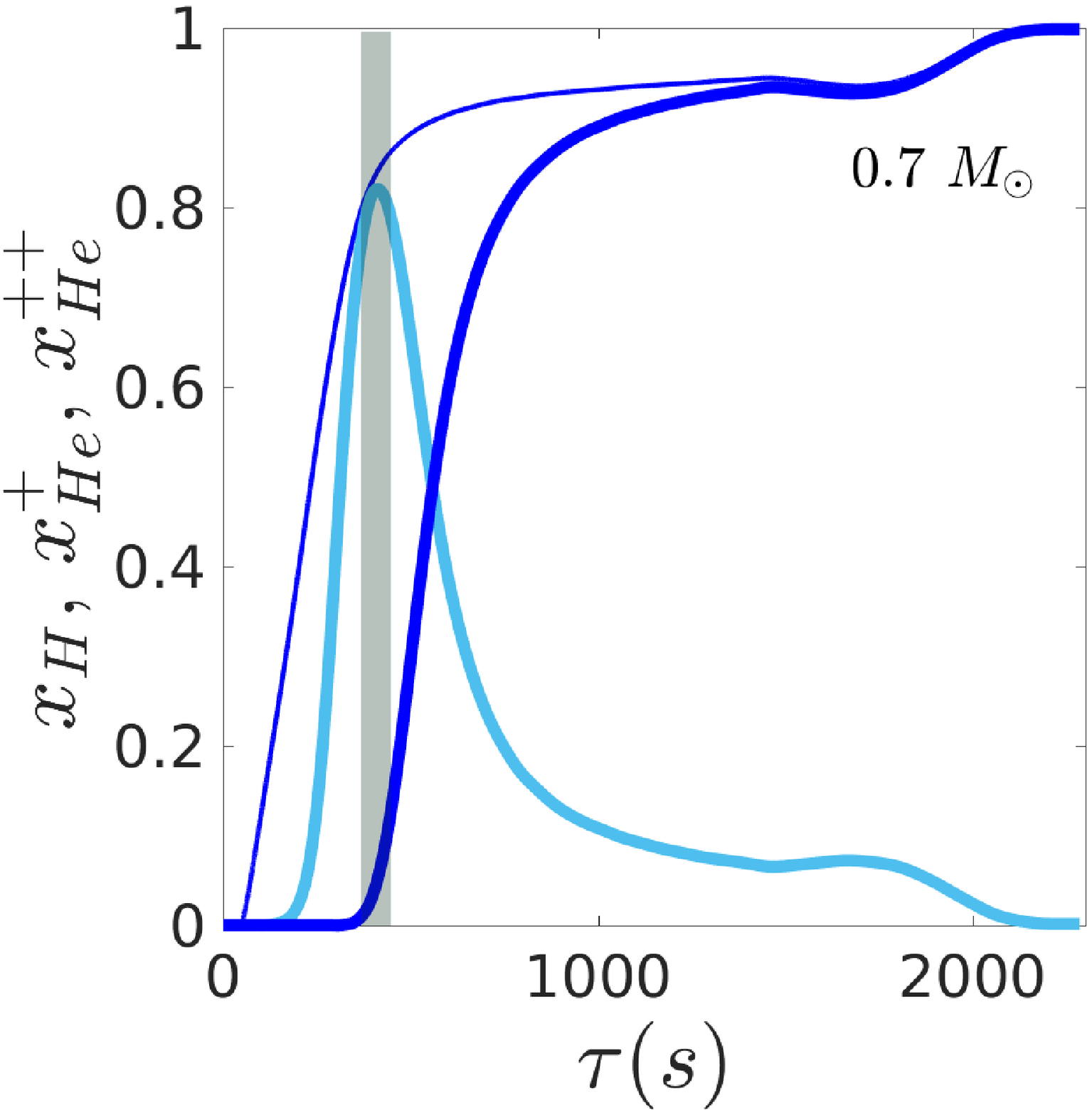}\\
	\includegraphics[width=0.30\textwidth]{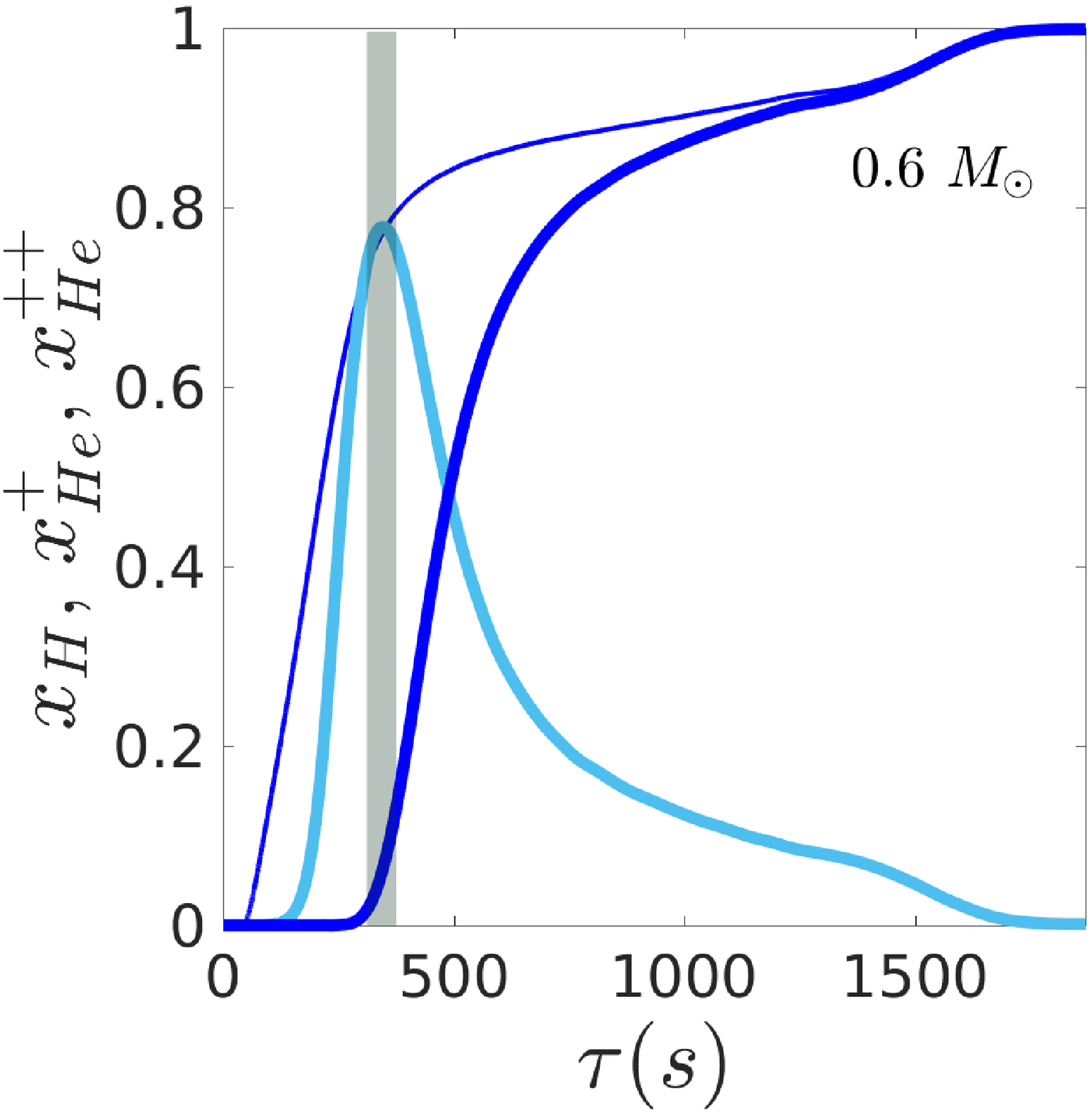}
	\includegraphics[width=0.30\textwidth]{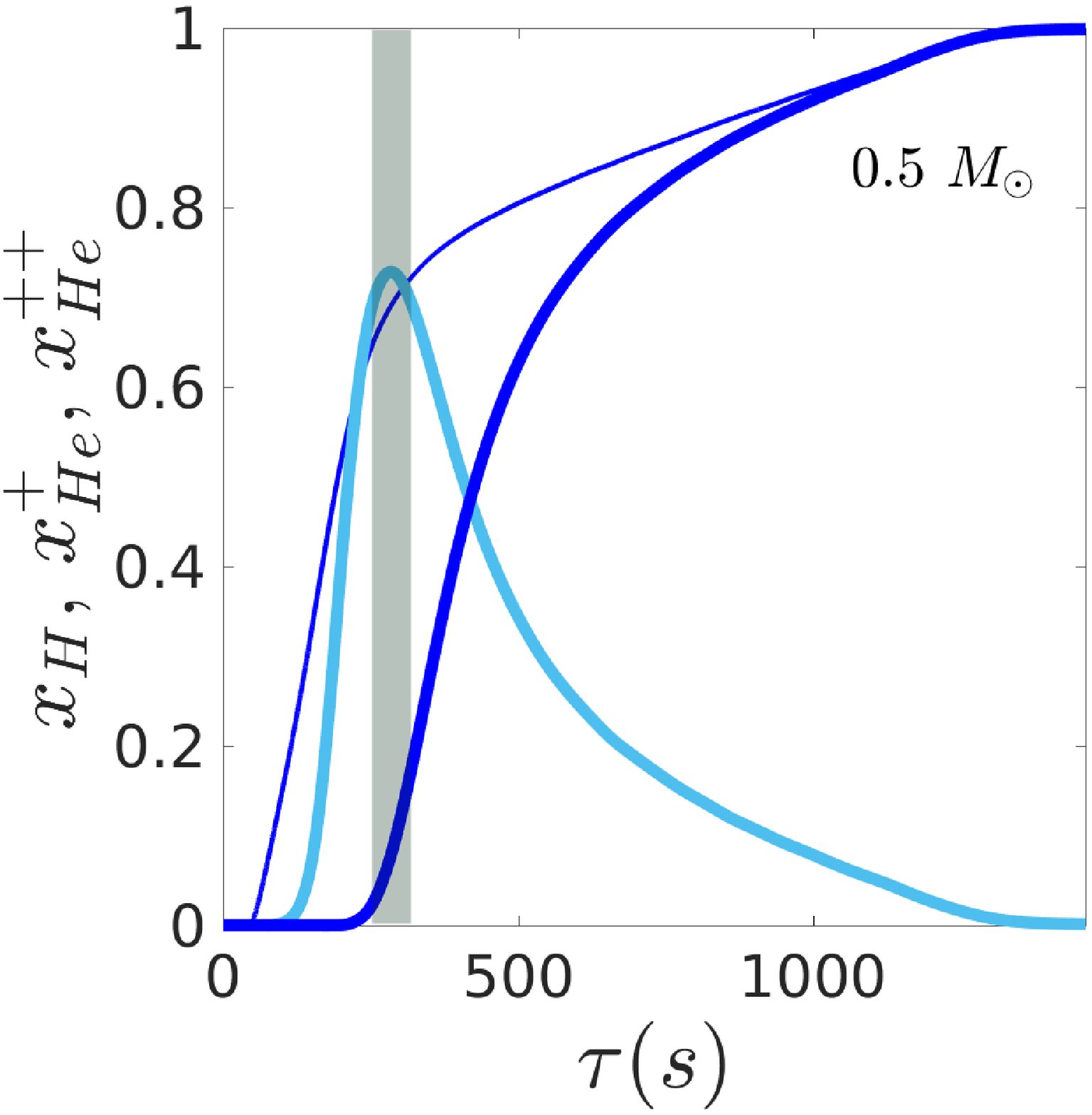}
	\includegraphics[width=0.30\textwidth]{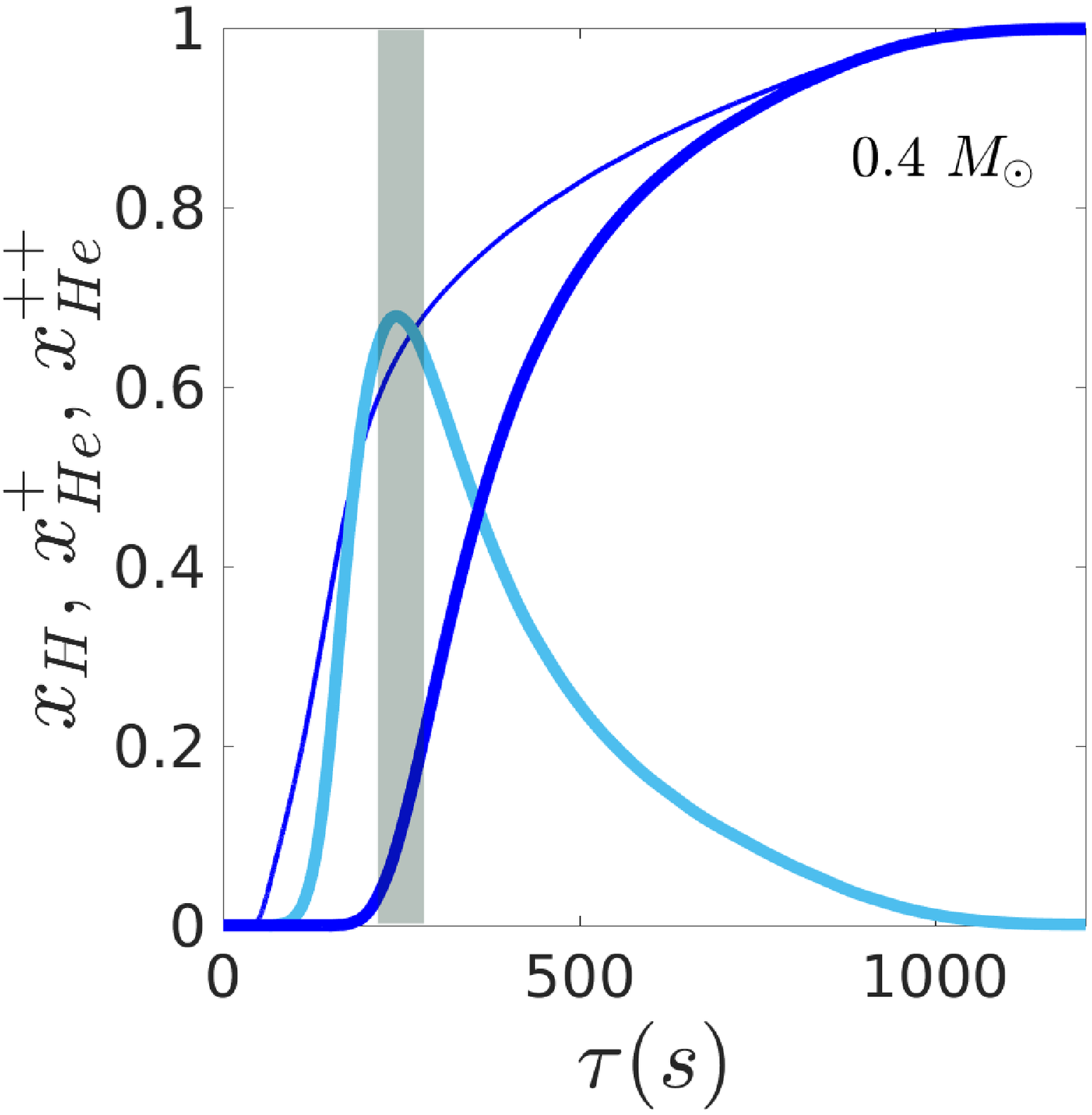}
	\caption{Ionisation fractions of hydrogen and helium for all the theoretical models. The vertical grey bar signalises the region where single ionised helium reaches its maximum value.}
	\label{fig6}
\end{figure*}

The stellar structure and composition determines how stars oscillate. From the equations of stellar oscillations, two characteristic frequencies emerge \citep[e.g.][]{2010aste.book.....A}. One is the acoustic, degree-dependent Lamb frequency, $S_l$, such that
\begin{equation}
	S_l^2 = \frac{l(l+1)}{r^2} c^2(r) \, .
\end{equation}
The Lamb frequency depends on the adiabatic sound speed $c(r)$ (eq. \ref{eq4}) and consequently also depends on the first adiabatic exponent $\Gamma_1$ (eq. \ref{eq3}). The other frequency is the  Brunt--V\"{a}is\"{a}l\"{a} (or buoyancy) frequency, $N$, which squared value is given by
\begin{equation}
	N^2 = g \left(\frac{1}{\Gamma_1} \frac{d \ln P}{dr} - \frac{d \ln \rho}{dr}\right) \, .
\end{equation}
These two characteristic frequencies allow us to identify two classes of solutions to the equations of stellar oscillations. Acoustic modes of oscillation (p--modes) have frequencies higher than both $S_l$ and $N$, whereas gravity modes (g--modes) have frequencies lower than both $S_l$ and $N$. Propagation diagrams are useful tools to represent the cavities where the different types of modes are trapped. The propagation diagrams for our theoretical models are shown in figure \ref{fig5}, where the frequency of maximum power, $\nu_{\text{max}}$, is represented by a solid horizontal grey line. This frequency is related to the near-surface properties of the star \citep{1991ApJ...368..599B, 1995A&A...293...87K},  $\nu_{\text{max}} \propto g \, T_{\text{eff}}^{-1/2}$, where $g$ is the surface gravity and $T_{\text{eff}}$ the effective temperature of the star. This scaling relation is used by MESA to determine the $\nu_{\text{max}}$ values for the stellar models. From an observational point of view, $\nu_{\text{max}}$ is the frequency where the power of the acoustic modes is strongest and its vicinity determines the window where is expectable to observe acoustic modes of oscillation. These diagrams highlight the clear acoustic character of the oscillations in the vicinity of $\nu_{\text{max}}$. Higher frequencies have a natural maximum which is known as the acoustic cutoff frequency ($\omega_c = c/2H$, where $H$ is the density scale height). For sound waves with frequencies higher than the cutoff frequency, the modes propagate outward through the atmospheres of the stars and cannot be considered completely trapped inside the star. Furthermore, the trapping regions of acoustic oscillations become shallower and narrower with the increase of the mode degree $l$. These properties of p modes separate their acoustic spectra from the higher frequencies and also from the higher degrees. Moreover, figure \ref{fig5} shows that in the case of the lowest mass models there is an upturn of the Lamb frequencies at the surface, turning the trapping regions of these stars particularly shallow and narrow.

As the mass of the stars decreases and the depth of convective zones increases, the g--mode cavity becomes shallower and the two distinct propagation regions move away from each other. Our study will focus on the outer convective regions where the most abundant elements undergo partial ionisation and the electrostatic interactions between particles have greater relevance.  
It is well known that the thermodynamic properties of stellar interiors depend on the degree of ionisation of the stellar matter. However, understanding partial ionisation zones is not a simple task since it requires to solve a system of interconnected equations \citep{1921RSPSA..99..135S}. Moreover, for low-mass main-sequence stars, other physical effects, such as Coulomb interactions,  can thwart the ideal gas approximation and should be taken into account \citep[e.g.][]{1977ApJS...35..293F}.

The most important partial ionisation zone for all the six stellar models is a region located approximately 2--4 per cent under the surface where light elements, hydrogen and helium, undergo partial ionisation (2\% for the more massive model, $0.9 \, M_\odot$, and 4\% for the less massive model, $0.4 \, M_\odot$). The ionisation fractions for hydrogen and helium as a function of the acoustic depth are shown in figure \ref{fig6}. We see that for more massive models the partial ionisation zone of hydrogen and the second partial ionisation zone of helium are separated as in the case of solar models \citep[][]{2012sse..book.....K}, whereas for less massive models these two partial ionisation zones overlap each other. With a vertical grey bar we highlighted, in figure \ref{fig6}, the region where the single ionised fraction of helium, $x^+_{He}$, reaches its maximum value. This vertical bar is also shown in figure \ref{fig2} where we can identify the maximum value of $x^+_{He}$ with the peak that separates the \ion{He}{i} from the \ion{He}{ii} ionisation zones for the three more massive models (top panel, figure \ref{fig2}). For the less massive models, the $\Gamma_1$ dip becomes shallow and it is impossible to identify this peak. The fact that the $\Gamma_1$ dip becomes shallow seems to be related to the overlap of the two helium ionisations. This was reported also by \citet{2014ApJ...794..114V} for stellar models in the mass range of $0.8 - 1.5 \, M_\odot$. Therefore, for stellar models with masses close to $0.6 \, M_\odot$, the $\Gamma_1$ peak vanishes. This has implications for seismology because the abrupt variation of the sound speed in helium ionisation zones will no longer leave a signature on the oscillation frequencies. 
\begin{figure*} 
	\centering
	\includegraphics[width=0.60\textwidth]{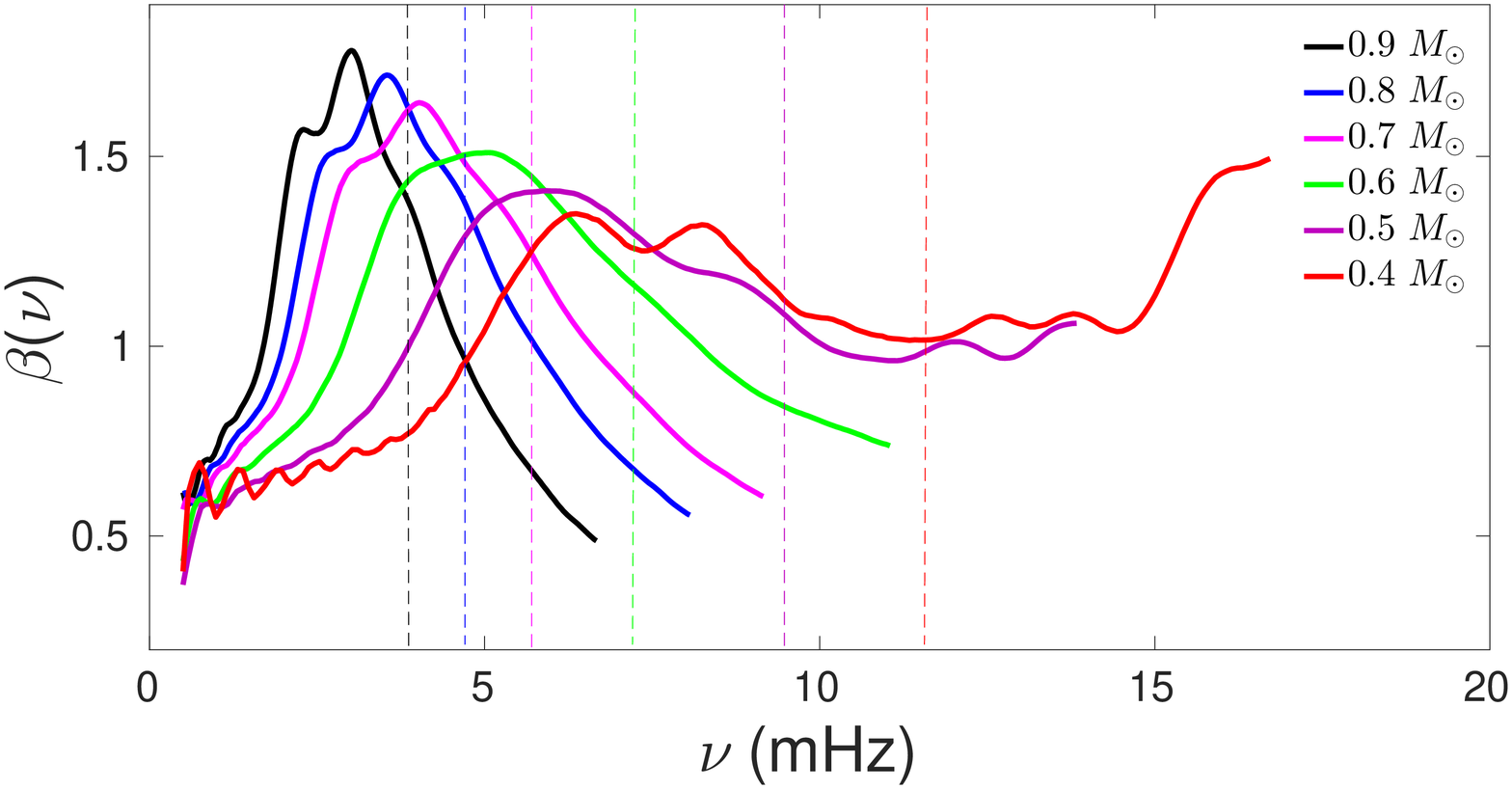}
	\includegraphics[width=0.30\textwidth]{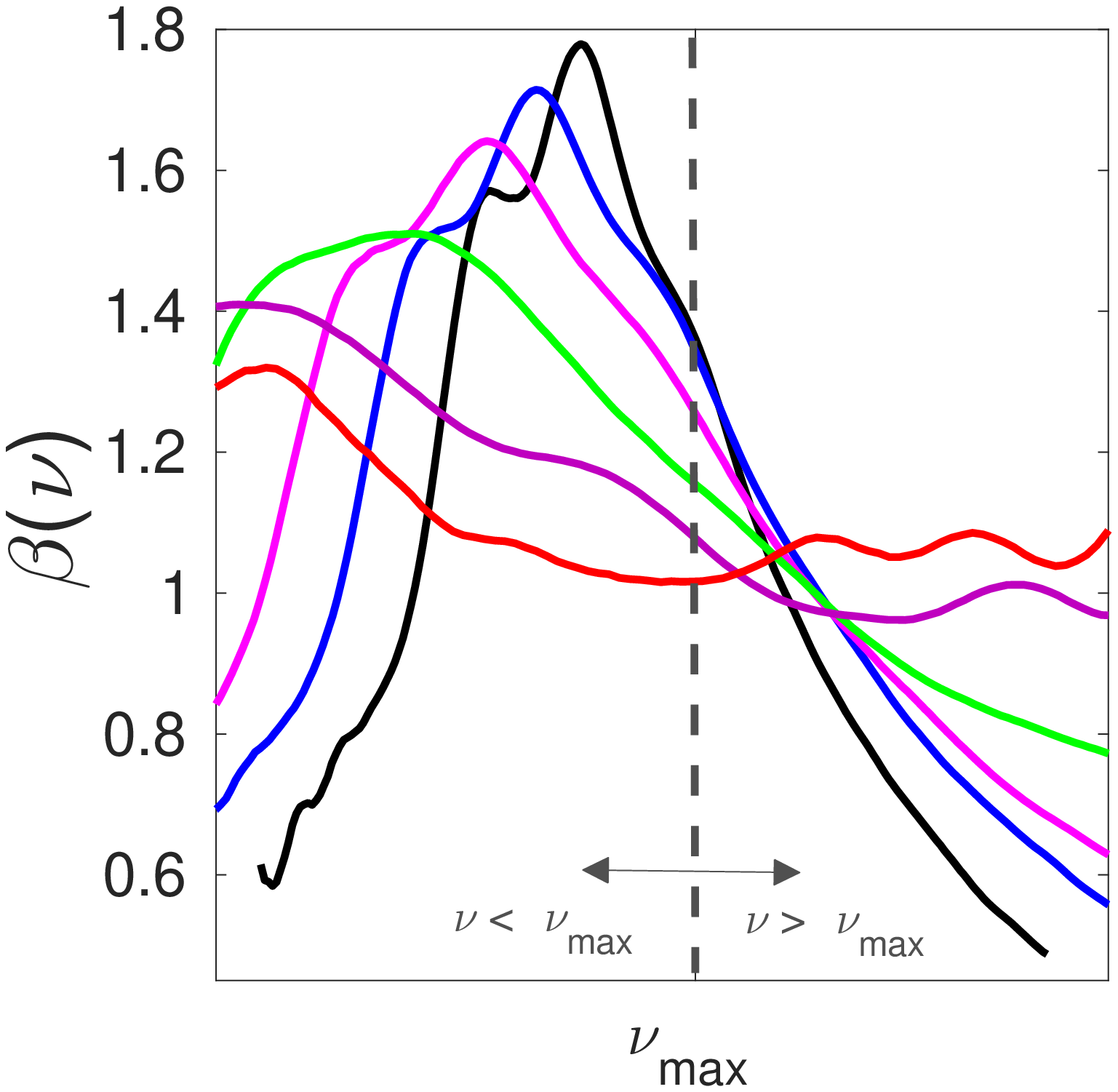}
	\caption{The seismic diagnostic $\beta(\nu)$ computed numerically from the structural model parameters for all the six models. Vertical dashed lines indicate the location of $\nu_{\text{max}}$. The right-hand panel shows all the diagnostics $\beta$ shifted and collapsed around the $\nu_{\text{max}}$ value.}
	\label{fig7}
\end{figure*}

\begin{figure*} 
	\centering
	\includegraphics[width=0.42\textwidth]{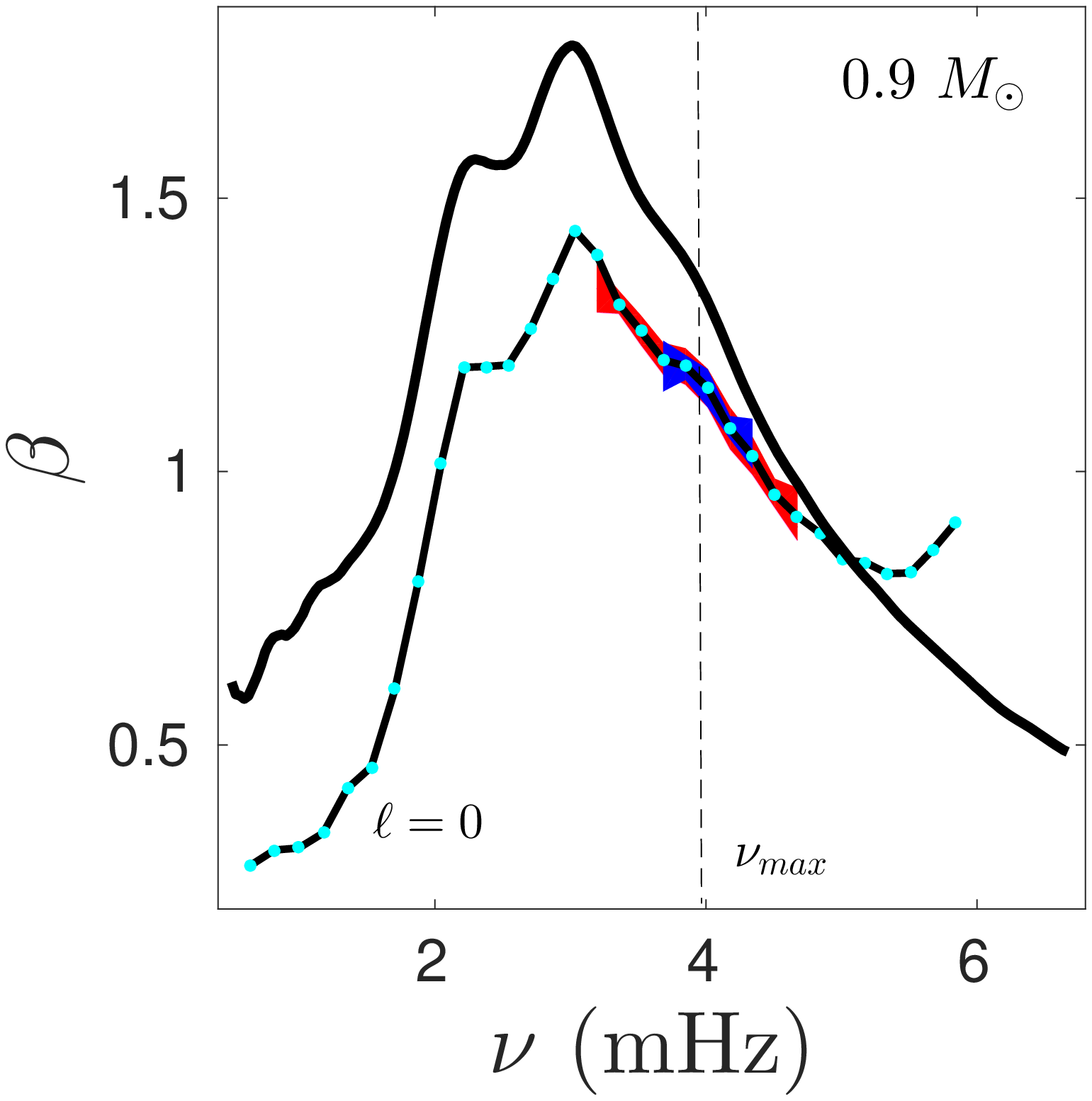}
	\includegraphics[width=0.42\textwidth]{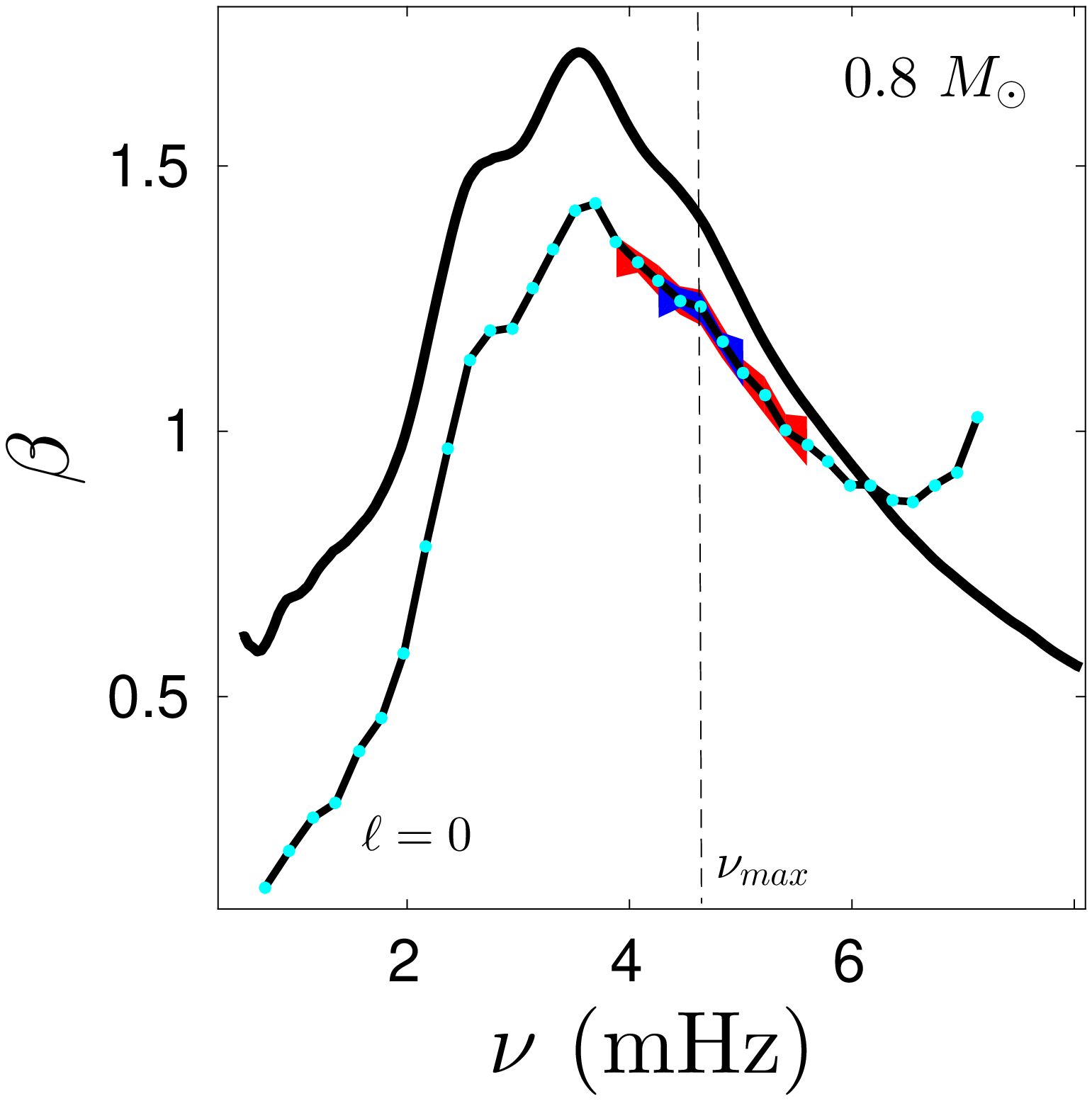}\\
	\includegraphics[width=0.42\textwidth]{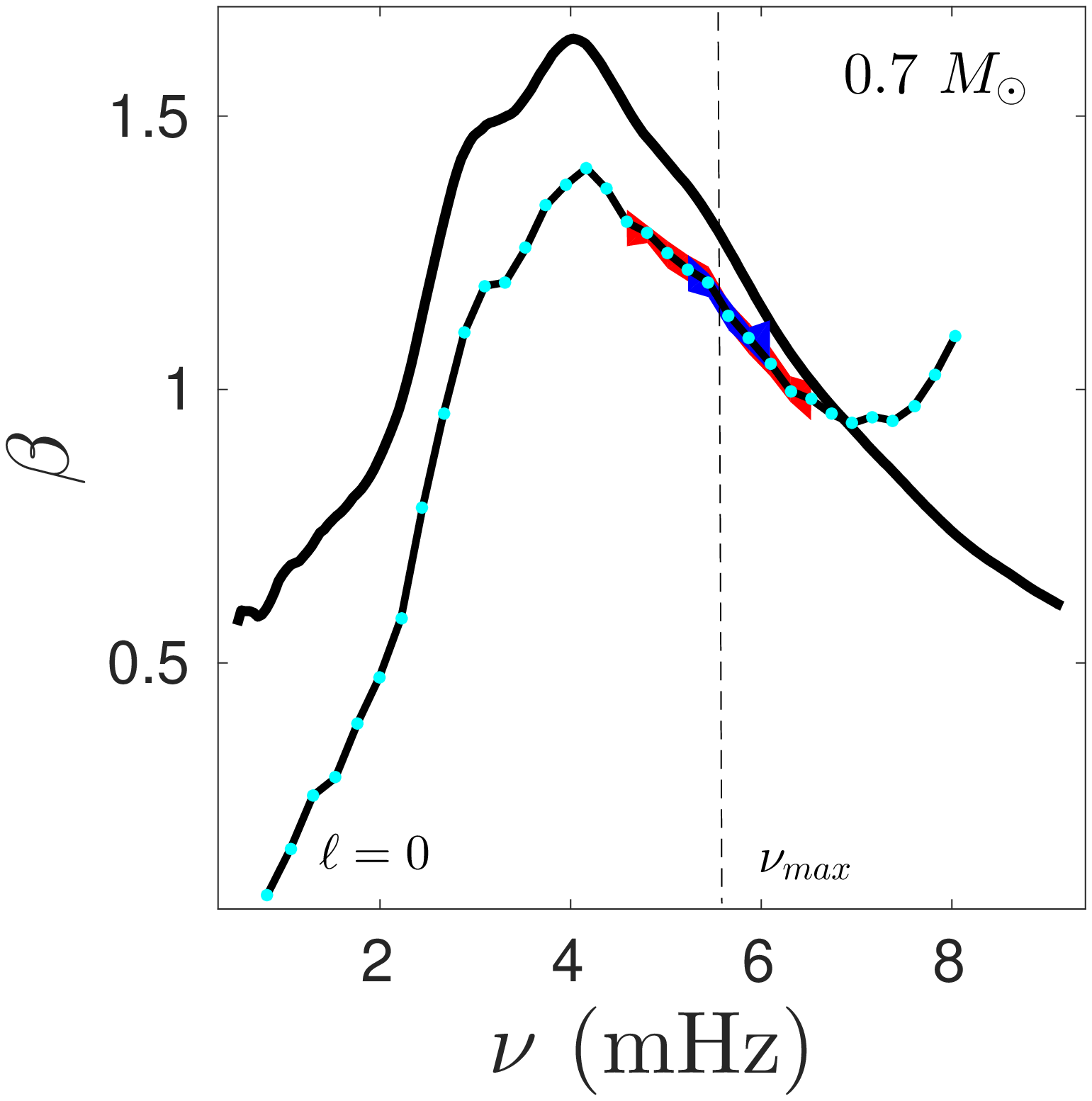}
	\includegraphics[width=0.42\textwidth]{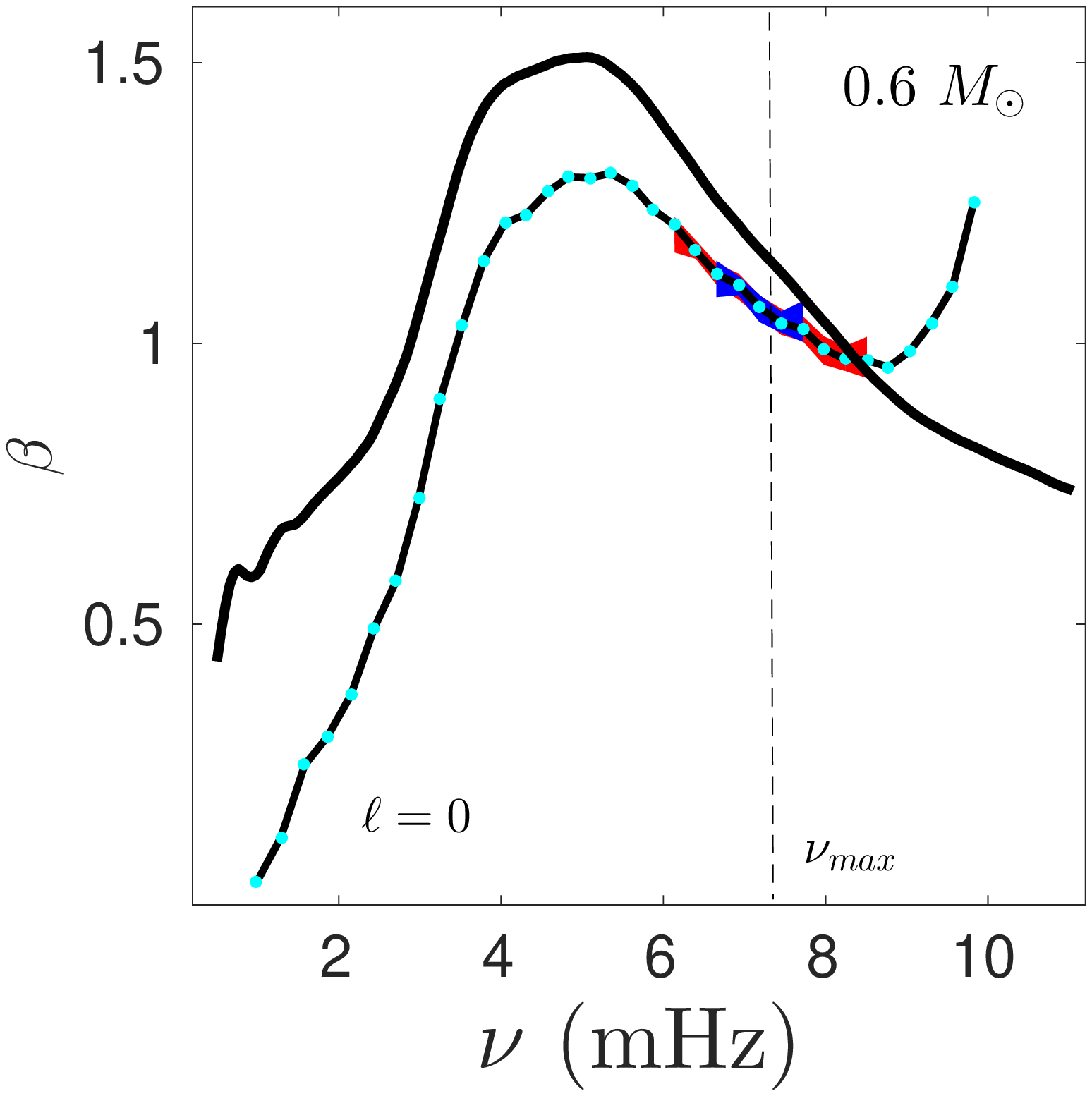}\\
	\includegraphics[width=0.42\textwidth]{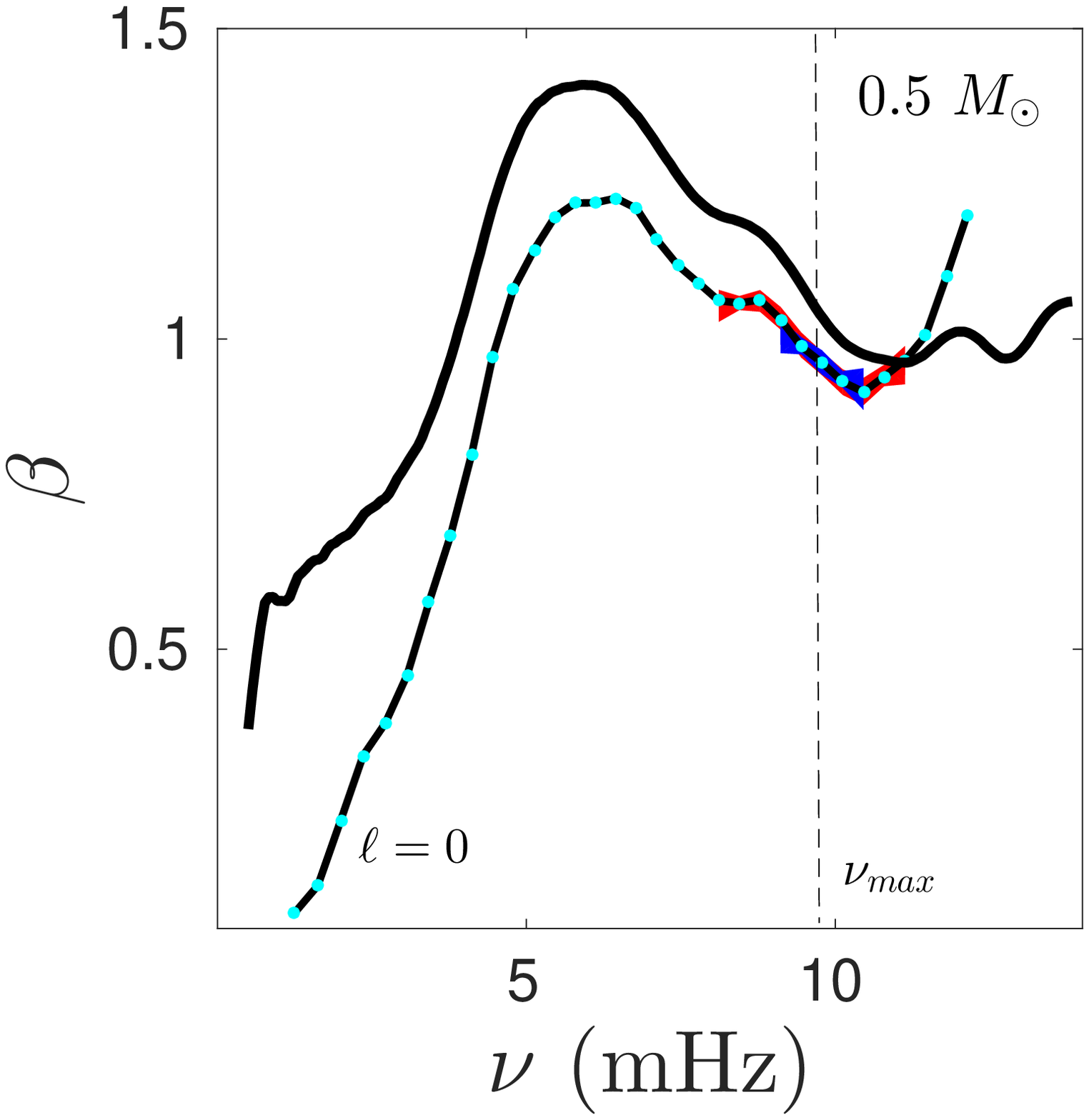}
	\includegraphics[width=0.42\textwidth]{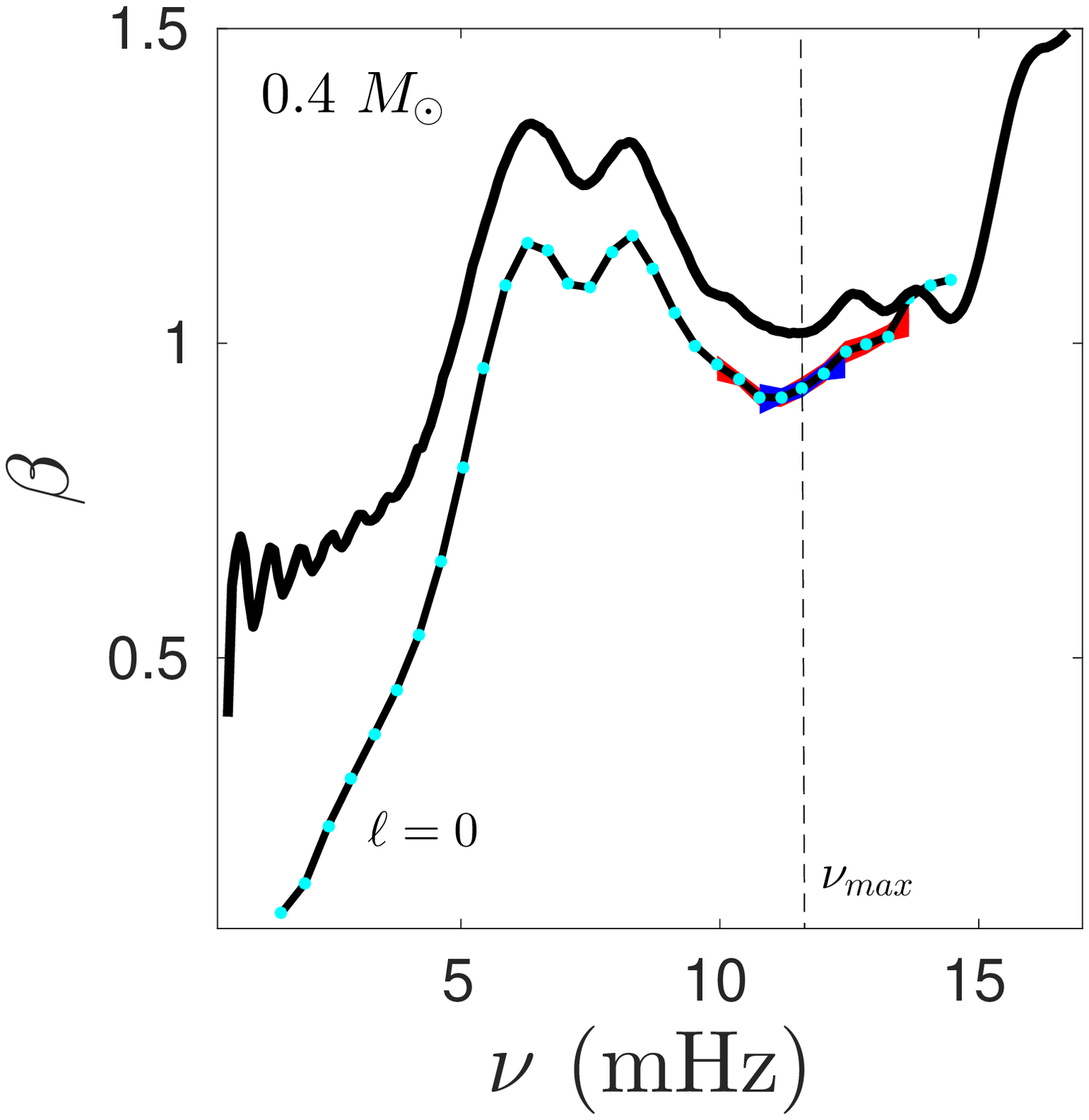}	
	\caption{Solid thick curves represent the seismic diagnostic $\beta(\nu)$ computed numerically from the structural model parameters for all the six models. Solid thin lines represent $\beta(\nu)$ computed from a theoretical table of frequencies for $l=0$. The case $l=0$ is highlighted with black colour and cyan markers. The shadowed regions were obtained by adding white noise ($\nu' = \nu + \Delta \nu$, where $\Delta \nu = 0.2$ $\mu$Hz) to a shorter selection of frequencies centred around $\nu_{\text{max}}$. Two cases were considered: (1) a table of frequencies centred around $\nu_{\text{max}}$ with 10 modes per degree -- red shadowed region; and (2) a table of frequencies centred around $\nu_{\text{max}}$ with 5 modes per degree -- blue shadowed region. The shadowed regions represented correspond to radial modes of oscillations. Results for $l=1, 2$  are similar but not shown, in order not to overload the figure.}
	\label{fig8}
\end{figure*}

Regions in stellar interiors where the sound speed undergoes an abrupt variation have been studied for several years in solar-type stars \citep[e.g.][]{2014ApJ...782...18M,  2019LRSP...16....4G, 2019MNRAS.489.1850V}. The abrupt variation of the sound speed produces a quasi-periodic oscillatory component in the frequencies of stellar oscillations \citep[e.g.][]{1988IAUS..123..151V, 1990LNP...367..283G}. The properties of this periodic signal contain important information about the location, specifically about the acoustic depth of the location that originated the signal  \citep[e.g.][]{2007MNRAS.375..861H, 2017ApJ...837...47V}.

To describe the characteristics of this signal, we will use a seismic diagnostic, $\beta(\nu)$, where $\nu$ stands for the cyclic frequency, which is particularly suitable to probe the outer convective layers of main-sequence stars \citep[e.g.][]{1989SvAL...15...27B, 1989ASPRv...7....1V, 1994A&A...290..845L, 2001MNRAS.322..473L}. Moreover, this seismic diagnostic is tightly related to the physical processes taking place in the convective zone \citep[e.g.][]{1997ApJ...480..794L}. This is so because the acoustic potential (eq. \ref{eq7}) allows us to determine, numerically, the acoustic outer phase shift of the reflected sound waves near the surface. The seismic diagnostic $\beta(\nu)$, in turn, is given by the derivative of this phase shift,
\begin{equation}\label{eq11}
	\beta(\nu) = - \nu^2 \frac{d}{d\nu} \left( \frac{\alpha}{\nu} \right) \, ,
\end{equation}
and can also be expressed as a function of the oscillation frequencies,
\begin{equation}\label{eq12}
	\beta(\nu) = \frac{\nu - n \left( \frac{\partial \nu}{\partial n} \right) - L \left( \frac{\partial \nu}{\partial L}  \right)}{ \frac{\partial \nu}{\partial n}} \, .
\end{equation}
In the above equation, $L = l+ 1/2$, and $l$ and $n$ represent, respectively, the degree of the acoustic mode and its radial order.
Figure \ref{fig7} shows the diagnostic $\beta(\nu)$ computed numerically using equation \ref{eq11} for all the models. We can clearly see how $\beta$ responds to modifications in the stellar structure. In particular, how the quasi-sinusoidal signal, quite marked for the $0.9 \, M_\odot$ model, slowly vanishes with decreasing mass. For the $0.6 \, M_\odot$ model, the signal is almost absent in the diagnostic $\beta$, as it is also absent the corresponding dip in the acoustic potential. For models with masses lower than $0.6 \, M_\odot$, the diagnostic $\beta$ reflects the complex pattern of perturbations in the acoustic potential. 

The frequencies around $\nu_{\text{max}}$ are particularly interesting because, as it is well known, it is around this value that is more likely to detect acoustic oscillation modes. Therefore, to highlight these frequencies, and also to better visualise the changes that occur in the $\beta(\nu)$ diagnostic while varying the mass of the models, we shifted all curves so that all $\nu_{\text{max}}$ frequencies will coincide. Considering a tangent line approximation at $\nu_{\text{max}}$ we note that as the impact of Coulomb effects increase (stellar mass decreases) the line presents a change in slope from negative to positive allowing us to discriminate between different cases.
The right panel of figure \ref{fig7}, which shows all the diagnostics $\beta(\nu)$ shifted to collapse around the frequency of maximum power, reinforces the idea that an important internal structural transition occurs around the models with $0.6 \, M_\odot$. Being a proxy for the surface phase shift, $\beta(\nu)$ depends on the complex structure of the outer layers of the star. 
High frequency modes (e.g. $\nu > \nu_{\text{max}}$) have upper turning points, which are located closer to the surface than modes with lower frequencies (e.g. $\nu < \nu_{\text{max}}$). Figure \ref{fig7} suggests that alongside with the scattering location located approximately 2 per cent under the surface of the star (represented by a vertical grey bar in figure \ref{fig2}), another relevant scattering location is located above the helium partial ionisation zone for stars less massive than $0.6 \, M_\odot$.
This different type of oscillatory signal present in the $\beta$ diagnostic of less massive stars contains valuable information about electrostatic interactions between the atomic species that compose the models. Here, heavy elements will certainly play an important role with some zones being more relevant than others from the point of view of electrostatic interactions; in the same way, some partial ionisation zones are more relevant than others. We will leave the study of these Coulomb zones for a future work.

For obtaining the theoretical p--mode adiabatic oscillation frequencies, we used the GYRE stellar oscillation code \citep{2013MNRAS.435.3406T, 2018MNRAS.475..879T, 2020ApJ...899..116G}. The individual mode frequencies were computed for degrees $l=0, 1, 2$ below the acoustic cutoff frequency. For each model, we obtained a set of 36 frequencies with radial orders in the range  $3 \le n \le 38$. The results of the seismic diagnostic $\beta(\nu)$ computed from a theoretical table of frequencies (eq. \ref{eq12}) were compared with $\beta(\nu)$ obtained numerically from the structural model parameters in figure \ref{fig8}. The discrepancy between the two results is a well-known effect related to the fact that the structural model parameters are obtained within the Cowling approximation that neglects the effects of the gravitational potential \citep[e.g.][]{2018ApJ...853..183B}. As in figure \ref{fig7}, it is clearly seen the transition from a signature dominated by the partial ionisation of helium to a signal dominated by electrostatic interactions between particles.

From an observational point of view, the acoustic oscillation spectrum of solar-like oscillators are well represented by a Gaussian-like envelope centred on the frequency of the maximum power. Therefore, a typical observational table of mode frequencies consists of a few modes per degree centred on $\nu_{\text{max}}$. To analyse a more realistic range of overtones, we considered two shorter frequency tables, i.e. two subtables of the full table mentioned above. One with 10 modes per degree and the other with 5 modes per degree, both centred on $\nu_{\text{max}}$. The outcome of any numerical differentiation strongly relies on the structure of the data. To validate the stability of the algorithm, we used these two tables and performed a series of tests. The tests consist of repeatedly (200 times) adding white noise to the frequencies, $\nu' = \nu + \Delta \nu$, of the two subtables of frequencies. The white noise was added within $0.2 \, \mu$Hz, which is twice the value of the uncertainty for the best Kepler targets \citep[e.g.][]{2017ApJ...835..172L}. The results are shown in figure \ref{fig8} for the case $l=0$ and validate the stability of the algorithm. Similar results were obtained for $l=1$ and $l=2$. The diagnostic potential of $\beta(\nu)$ increases considerably with the number of available modes to use in its computation. Nevertheless, even in the more restricted case of a frequency table with five modes per degree $l$, the behaviour of $\beta$ for a $0.9 \, M_\odot$ stellar model  is clearly distinct from that of $0.4 \, M_\odot$ as shown in figure \ref{fig8}. We also stress that even in this more restricted case, the $\beta$ curve preserves the properties and characteristics of the curve obtained with the full theoretical data table. 

\section{Conclusions} \label{sec4}

We perform a theoretical study of the outer convective layers of low-mass stars with masses in the range $0.4 - 0.9 \, M_\odot$. This is accomplished with the help of a powerful seismic diagnostic that is particularly sensitive to the outer convective layers of low-mass stars. Specifically, we study the oscillatory signature imprinted in the oscillation mode frequencies as a consequence of two distinct, non-ideal, physical processes: the partial ionisation of chemical elements and the electrostatic interactions between particles in the gas.

We find that, around the value of $0.6 \, M_\odot$, the peak/dip in the first adiabatic exponent/acoustic potential corresponding to helium ionisation zones (fig. \ref{fig2}) vanishes and, consequently, does not produce an oscillatory signature in the seismic diagnostic $\beta(\nu)$. This value, $0.6 \, M_\odot$, is thus the reference value, for models with solar metallicity, around which occurs a transition in the characteristics of the oscillatory signal imprinted in the oscillation frequencies by two distinct physical processes. The acoustic spectrum of stars more massive than $0.6 \, M_\odot$ is mostly impacted by partial ionisation processes, whereas the acoustic spectrum of stars less massive than $0.6 \, M_\odot$ is predominantly impacted by Coulomb interactions.

Several observational and theoretical studies with low-mass stars indicate that rotation and magnetic fields strongly influence stellar evolution \citep[e.g.][]{1984ApJ...279..763N, 2003ApJ...586..464B, 2003A&A...397..147P, 2008ApJ...687.1264M, 2009IAUS..258..363I, 2014ApJS..211...24M, 2014ApJ...794..144R, 2015MNRAS.450.1787A, 2018A&A...616A.108B, 2020arXiv200907695J}. Among cool stars with convective envelopes, it is known that hotter stars are generally faster rotators than cooler stars. This is thought to be related to the process of angular momentum loss caused by magnetic stellar winds \citep[e.g.][]{1962AnAp...25...18S, 1972ApJ...171..565S, 1988ApJ...333..236K}. Furthermore, the magnetic field that originates the stellar magnetised wind is amplified by a dynamo mechanism at the transition between the convective envelope and the radiative interior \citep[e.g.][]{2017LRSP...14....4B}. Therefore, it seems that the magnetic and rotational properties of stars are closely related to specific aspects of the stellar structure, and in the particular case of G, K, and early M stars, the dynamics and the underlying microphysics of stellar convective zones are of special importance. Recent results, such as evidence for a broken spin down in K2 data \citep{2021arXiv210107886G} or a reported metallicity dependence on stellar rotation \citep{2020MNRAS.499.3481A, 2021arXiv210305675S}, are interesting results that allow us to reinforce the importance of observing low-mass stars with convective envelopes aiming in obtaining asteroseismic data. Our work unveils the potential that future discoveries related to the detection of solar-like oscillations in stars less massive than the Sun could bring to the study of the relationships between the internal structure of low-mass stars and their observable characteristics.

We expect that in the future, space missions like PLATO \citep[e.g.][]{2014ExA....38..249R} or the ground-based promising observations of cool stars like SONG \citep{2017ApJ...836..142G} or KPF  \citep{2016SPIE.9908E..70G}, could provide enough observational data to put constraints on the internal structure of these stars and significantly help to improve stellar evolution models of low-mass stars.

\section*{Acknowledgements}

We thank the anonymous referee for all the comments and suggestions that helped us significantly to improve the manuscript.
We are also grateful to Bill Paxton/Rich Townsend and to all the development team members for the great work that makes possible the public use of the stellar evolution code MESA/stellar oscillation code GYRE. 
The authors A. Brito and I. Lopes thank the Funda\c c\~ao para a Ci\^encia e Tecnologia (FCT), Portugal for the financial support to the Center for Astrophysics and Gravitation-CENTRA, Instituto Superior Tecnico, Universidade de Lisboa, through the Project No. UIDB/00099/2020. and grant No. PTDC/FISAST/ 28920/2017.

\section*{Data Availability Statement}
This work makes use of the 1D stellar evolution code MESA (Modules for Experiments in Stellar Astrophysics) publicly available at the Zenodo repository under the DOI: 10.5281/zenodo.3473377. The essential files to reproduce the results are also available at Zenodo with the DOI: 10.5281/zenodo.4646955. The GYRE stellar oscillation code was used to calculate the eigenfrequencies for each stellar model. GYRE is also publicly available at https://gyre.readthedocs.io/en/stable/.




\bibliographystyle{mnras}
\bibliography{p6} 




%
%


\bsp	
\label{lastpage}
\end{document}